\documentclass[pre,longbibliography,nofootinbib,tighten]{revtex4-2}

\usepackage[utf8]{inputenc}
\usepackage{amssymb,amsmath}
\usepackage{epsfig}
\usepackage{bm}
\usepackage{color}

\usepackage{graphicx}

\usepackage{color}

\newcommand\beq{\begin{equation}}
\newcommand\eeq{\end{equation}}
\newcommand\beqa{\begin{eqnarray}}
\newcommand\eeqa{\end{eqnarray}}
\newcommand{\dd}{\text{d}}

\newcommand{\al}{\alpha}

\begin{document}

\title{Nonlinear transport of tracer particles immersed in a strongly sheared dilute gas with inelastic collisions}

\author{David Gonz\'alez M\'endez\footnote[1]{Electronic address: dgonzalezm@unex.es}}
\affiliation{Departamento de F\'{\i}sica,
Universidad de Extremadura, E-06006 Badajoz, Spain}
\author{Vicente Garz\'o\footnote[2]{Electronic address: vicenteg@unex.es;
URL: https://fisteor.cms.unex.es/investigadores/vicente-garzo-puertos/}}
\affiliation{Departamento de F\'{\i}sica and Instituto de Computaci\'on Cient\'{\i}fica Avanzada (ICCAEx), Universidad de Extremadura, E-06006 Badajoz, Spain}

\begin{abstract}

Nonlinear transport of tracer particles immersed in a sheared dilute gas with inelastic collisions is analyzed within the framework of the Boltzmann kinetic equation.  Two different yet complementary approaches are employed to obtain exact results. First, we maintain the structure of the \textit{inelastic} Boltzmann collision operator but consider inelastic Maxwell models (IMM) instead of the realistic model of inelastic hard spheres (IHS). Using IMM enables us to compute the collisional moments of the inelastic Boltzmann operator for mixtures without explicitly knowing the velocity distribution functions of the mixture. Second, we consider a kinetic model of the Boltzmann equation for IHS. This kinetic model is based on the equivalence between a gas of elastic hard spheres subjected to a drag force proportional to the particle velocity and a gas of IHS. We solve the Boltzmann--Lorentz kinetic equation for tracer particles using a generalized Chapman--Enskog--like expansion around the shear flow distribution. This reference distribution retains all hydrodynamic orders in the shear rate. The mass flux is obtained to first order in the deviations of the concentration, pressure, and temperature from their values in the reference state. Due to the anisotropy induced in the velocity space by shear flow, the mass flux is expressed in terms of tensorial quantities rather than conventional scalar diffusion coefficients. Unlike the previous results obtained for IHS using different approximations, the results derived in this paper are exact. Generally, the comparison between the IHS results and those found here shows reasonable quantitative agreement, especially for IMM results. This good agreement shows again evidence of the reliability of IMM for studying rapid granular flows. Finally, we analyze segregation by thermal diffusion as an application of the theory. Phase diagrams illustrating segregation are presented and compared with previous IHS results, demonstrating qualitative agreement.

\end{abstract}

\maketitle

\section{Introduction}
\label{sec1}

Granular materials are collections of solid particles that interact primarily through contact. Examples in industry include grains, rice, gravel, powders, coffee beans, and pharmaceutical pills. They are also important in geological processes such as landslides, snow avalanches, erosion and/or plate tectonics. Granular materials
have a few things in common, one of which is that they have macroscopic dimensions; therefore, their interactions are dissipative. This feature likely makes granular flows distinct from conventional flows observed in molecular gases or liquids. When subjected to violent and sustained excitation, the motion of the grains resembles the random motion of atoms or molecules in a molecular gas. Under these conditions, known as rapid flow regime, the system admits a hydrodynamic-type description, and the granular material is often referred to in the literature as a \textit{granular} gas \cite{G03}.

To analyze the impact of inelasticity in collisions on the dynamical properties of the solid particles an idealized model is usually considered in the rapid flow regime. The idealized model is a gas of hard spheres with instantaneous inelastic collisions (inelastic hard spheres, IHS). In the simplest version of the model, the spheres are assumed to be completely smooth and so, the inelasticity of collisions is accounted for by a (positive) coefficient of restitution.

Under rapid flow conditions, the tools of classical kinetic theory of gases, adapted to dissipative dynamics, can be used to derive the corresponding hydrodynamic equations for the IHS model with explicit expressions for the transport coefficients \cite{C90,G03,BP04,RN08,G19}. In particular, for low-density gases, the
conventional Boltzmann equation has been extended to IHS \cite{BP04,G19} and the kinetic equation has been solved by means of the Chapman--Enskog method \cite{CC70} to first order in the spatial gradients. The corresponding Navier--Stokes transport coefficients are given in terms of the solution of a set of coupled linear integral equation. However, as with elastic collisions \cite{CC70,FK72}, for hard spheres usually these integral equations are \emph{approximately} solved by considering the first few terms in a Sonine polynomial expansion of the distribution function. This procedure can be extended to the more realistic case of mixtures, namely when grains have different masses and/or diameters. However, the problem is more complex than that of a monocomponent gas since the number of transport coefficients is greater and they depend on more parameters (see, for example, Refs.\ \cite{JM87,JM89,Z95,GD02,SGNT06}). Additionally, determining the transport properties from the Boltzmann equation for both elastic and/or inelastic hard spheres is a very difficult task for far from equilibrium states (i.e., beyond the Navier--Stokes domain).

Due to the aforementioned difficulties when the hard sphere kernel is employed, alternative approaches are commonly used in kinetic theory to obtain exact results. One possibility is to maintain the intricate mathematical structure of the Boltzmann collision operator while assuming a different interaction model: the so-called inelastic Maxwell model (IMM). To the best of our knowledge, this interaction model was introduced independently in Refs.\ \cite{BK00} and \cite{BCG00} at the beginning of the 21st century. The main reason for introducing IMM was to analyze in a clean way the overpopulation associated with the high energy tails of the distribution function in the homogeneous cooling state (namely, a homogeneous state with a temperature decaying in time) \cite{BCG00,EB02b,EB02a,BMP02,KB02,KB02a,EB02c,BC02,BK03,BC03,BCT03,ETB06,BTE07}. As with the conventional Maxwell molecules \cite{E81}, the collision rate for IMM is independent of the relative velocity of the colliding spheres. This contrasts with the IHS model, in which the collision rate is proportional to the relative velocity. The main advantage of using IMM instead of IHS is that a collisional moment of degree $k$ can be expressed in terms of velocity moments of degree $k$ or smaller than $k$, without knowing the velocity distribution function. This property allows one to obtain the exact forms of the Navier-Stokes transport coefficients \cite{S03}, as well as the rheological properties of sheared granular gases \cite{G03bis}.

Another possible approach to achieving exact results is considering simpler kinetic models of the Boltzmann equation for IHS. These models are mathematically more tractable than the Boltzmann equation (since they usually replace the original Boltzmann collision operator by a simple relaxation term) and typically maintain the primary physical properties of the true kinetic equation. In particular, for elastic collisions, the well-known Bhatnagar--Gross--Krook (BGK) kinetic model \cite{BGK54,GS03} has proven to be an accurate tool to determine nonlinear transport properties, especially in shearing nonequilibrium states \cite{D90,GS03}. Several models have been proposed for inelastic collisions in single gases \cite{BMD96,BDS99,DBZ04}. However, the number of kinetic models proposed in the literature for multicomponent granular gases is much smaller. We are aware of only one kinetic model
reported in the granular literature: the model proposed by Vega Reyes \emph{et al.} (VGS model) \cite{VGS07}. This contrasts with the large number of kinetic models proposed in the case of molecular mixtures (see, for instance, Refs.\ \cite{GK56,S62,H65,H66,GS67,GSB89,AAP02,HHKPW21,LZW24}). The VGS model is based on the equivalence between a gas of elastic hard spheres subjected to a drag force proportional to the particle velocity and a gas of IHS \cite{SA05}. The kinetic model is defined in terms of a relaxation term that can be selected from among the various models proposed for molecular mixtures. Recently \cite{AGG25}, the VGS model has been solved in three different nonequilibrium problems. A comparison of the results derived in \cite{AGG25} with those obtained from the Boltzmann equation for IHS shows a general agreement, especially when dissipation is not strong.

One of the most interesting nonequilibrium problems in granular flows is the diffusion of tracer particles in a granular gas. Studying diffusion in a granular binary mixture where one of the species is present in tracer concentration is more amenable to analytical treatment than studying the general case, since the state of the excess species is unaffected by the presence of tracer particles. Most previous studies on this topic \cite{BP00a,BRCG00,DBL02,GM04} have been focused on tracer diffusion in a freely cooling granular gas (homogeneous cooling state, HCS). In this situation, analytical (approximate) expressions have been derived for the tracer diffusion coefficient for IHS \cite{BP00,BRCG00,DBL02,GM04} as well as for IMM \cite{G19}. These expressions have been shown to agree very well with numerical results obtained from the direct simulation Monte Carlo (DSMC) method \cite{B94}. Nevertheless, beyond the HCS, studies on tracer diffusion are scarce. In particular, at the level of kinetic theory  and within the context of the Boltzmann equation for IHS, some previous works \cite{G02,G07} have analyzed the diffusion of impurities in a granular gas under uniform shear flow (USF). As in the case of the HCS, the results obtained in Refs.\ \cite{G02,G07} are approximate since they were obtained by considering the leading term in a Sonine polynomial expansion of the distribution functions. Thus, it would be useful to revisit the previously analyzed problem for IHS \cite{G02,G07} but starting from a kinetic equation where exact results can be obtained.

The objective of this paper is to analyze the diffusion of tracer particles in sheared granular gases from the \emph{inelastic} Boltzmann equation. As previously mentioned, two different yet complementary approaches are taken. First, we consider IMM and so the collision rate of the two colliding particles appearing inside of the Boltzmann collision operators $J_{ij}[f_i,f_j]$ ($f_i$ and $f_j$ being the velocity distribution functions of species $i$ and $j$) is independent of the relative velocity. Using the Maxwell kernel instead of the hard sphere kernel enables us to precisely determine the moments of $J_{ij}[f_i,f_j]$ necessary for determining the nonlinear rheological properties in the USF. Thus, we can avoid using uncontrolled approximations to achieve analytical results, as in the case of IHS. Although the transport properties of USF can be in principle determined from IMM, the intricate mathematical structure of the Boltzmann operators prevents us the possibility from deriving the explicit forms of the velocity distribution functions. For this reason, and to complement the IMM results, we also consider the simple VGS kinetic model \cite{VGS07} as a second approach. Apart from obtaining the non-Newtonian transport properties of the system, the simplicity of the kinetic model allows us to explicitly get the velocity distribution functions in the USF. This is likely one of the main advantages of using a kinetic model instead of the true Boltzmann equation.

The study of tracer diffusion in granular shear flows has attracted the attention of engineers and physicists for years. Additionally, this is a problem of practical interest because granular materials must be mixed before processing can begin. Due to the complexity of the general problem,
the limiting situation where the tracer particles are mechanically equivalent to the particles of the granular gas (self-diffusion problem) was widely studied in earlier works. Thus, experimental studies include both systems with macroscopic flows \cite{NHT95,MD97} and vertical vibrated systems \cite{ZS91}. As a complement, computer simulation works \cite{C97,ZPTMS98,ALJR21} on dense systems have mainly analyzed the influence of the solid volume fraction on the elements of the self-diffusion tensor. As said before, analytical studies on this problem are scarce and to the best of our knowledge only the works \cite{G02,G07} have studied this problem. Unlike previous computational studies \cite{C97,ZPTMS98,ALJR21}, the analysis performed in Refs.\ \cite{G02,G07} considers tracer and granular gas particles as mechanically distinct, resulting in energy nonequipartition as the coefficients of restitution decrease. Even in the tracer limit case, studying mass transport under USF is quite intricate due to the cross-effects induced by shear flow that appear in mass transport. This gives rise to tensorial quantities ($D_{k\ell}$, $D_{p,k\ell}$, and $D_{T,k\ell}$) instead of the conventional scalar coefficients (the diffusion coefficient $D$, the pressure diffusion coefficient $D_p$, and the thermal diffusion coefficient $D_T$) for characterizing the mass transport.

Searching for exact solutions in kinetic theory is interesting from both a formal point of view and as a means of gauging the reliability of these types of solutions. Here, we compare the exact results
derived from IMM and the VGS kinetic model with those obtained approximately for IHS and in some cases with computer simulations available in the granular literature. Because the strength of the shear rate in the USF problem is arbitrary, this comparison can be considered a stringent test of the IMM and VGS models' ability to capture the trends observed in realistic granular flows.

One of the main limitations of using Maxwell models for granular gases is that these models do not describe real particles since they  do not interact according to a specific interaction potential. This contrasts with the situation for molecular gases, where elastic Maxwell models are consistent with a repulsive potential that is inversely proportional to the fourth power of the distance in three dimensions \cite{E81}. Nevertheless, many researchers working in the field acknowledge that the loss of physical realism can be offset by the quantity of exact analytical results obtained using IMM. As Ernst and Brito \cite{EB02b} claim ``what harmonic oscillators are for quantum mechanics and dumb-bells for polymer physics, is what elastic and inelastic Maxwell models are for kinetic theory.'' Apart from their mathematical tractability, it should be noted that some experiments involving magnetic grains with dipolar interactions have been well described by IMM \cite{KSSAB05}.

Regarding the use of kinetic models for IHS instead of the original Boltzmann equation, many previous studies on sheared molecular gases have clearly demonstrated the reliability of these models in quantitatively reproducing the Boltzmann results for the rheological properties of the gas (see for instance, the review \cite{GS03}). However, beyond the second-degree velocity moments, significant discrepancies for the fourth-degree moments obtained from the Boltzmann and BGK equations have been found in the USF for strong shear rates \cite{GS03}. In this context and based on a previous comparison with DSMC results for the velocity distribution function of the USF \cite{BRM97}, it is expected that the VGS kinetic model will provide accurate predictions for the distribution function at low velocities. However, significant differences may appear in the high-velocity region.

The plan of the paper is as follows. Section \ref{sec2} introduces the Boltzmann equation and its balance hydrodynamic equations, and presents the form of the Boltzmann collision operators for IMM and the VGS kinetic model. Section \ref{sec3} analyzes the rheology of a granular mixture under USF in the tracer limit within the context of IMM. Once the rheological properties are determined, the elements of the diffusion tensors are explicitly obtained in section \ref{sec4} for IMM by solving the Boltzmann equation by means of a generalized Chapman--Enskog expansion \cite{CC70} around the shear flow distribution. Section \ref{sec5} briefly shows how to evaluate the tracer diffusion coefficients using the VGS model, while the comparison between the results obtained for IHS, IMM and the VGS model is carried out in section \ref{sec6}. The comparison generally shows reasonable agreement, especially for IMM. As an application, section \ref{sec7} addresses the problem of segregation by thermal diffusion. For the sake of simplicity, we consider a situation in which the thermal gradient is perpendicular to the shear flow plane ($xy$-plane), so only segregation parallel to the thermal gradient occurs in the system. In this situation, the sign of the thermal diffusion factor $\Lambda_z$ characterizes the tendency of the tracer particles to move towards the hot or cold plate. The paper ends in section \ref{sec8} with a discussion of the results obtained in the paper.

\section{Boltzmann kinetic equation for granular mixtures}
\label{sec2}

We consider a granular binary mixture constituted by smooth hard disks ($d=2$) or spheres ($d=3$) of masses $m_i$ and diameters $\sigma_i$ ($i=1,2$). The collisions between grains are inelastic and characterized by the (constant) coefficients of normal restitution $\al_{ij}$ ($0\leq \al_{ij}\leq 1$). The inelasticity in collisions only affects the translational degrees of freedom of grains. At a kinetic theory level, all the relevant information on the state of the mixture is given through the knowledge of the one-particle velocity distribution $f_i(\mathbf{r}, \mathbf{v}; t)$ of species $i$. In the low-density regime and in the absence of external forces, the distributions $f_i$ obey the set of coupled nonlinear (inelastic) Boltzmann equations
\begin{equation}
\label{2.1}
\left(\partial_t+{\bf v}\cdot \nabla \right)f_{i}
({\bf r},{\bf v};t)
=\sum_{j}J_{ij}\left[{\bf v}|f_{i}(t),f_{j}(t)\right],
\end{equation}
where $J_{ij}\left[{\bf v}|f_{i},f_{j}\right]$ are the Boltzmann collision operators \cite{G19}. The most relevant hydrodynamic fields in a binary mixture are the number densities $n_i$, the mean flow velocity  $\mathbf{U}$, and the granular temperature $T$. In terms of the distributions $f_i$, those fields are defined  as
\begin{equation}
\label{2.2}
n_i=\int \mathrm{d}{\bf v} f_i({\bf v}),
\end{equation}
\begin{equation}
\label{2.3}
 \rho{\bf U}=\sum_{i=1}^2\rho_i{\bf U}_i=\sum_{i=1}^2\int \mathrm{d}{\bf v}m_i{\bf v}f_i({\bf v}),
\end{equation}
\begin{equation}
\label{2.4}
nT=p=\sum_{i=1}^2 n_iT_i=\sum_{i=1}^2\frac{m_i}{d}\int \mathrm{d}{\bf v}V^2f_i({\bf v}),
\end{equation}
where $\rho_i=m_in_i$ is the mass density of species $i$, $n=n_1+n_2$ is the total number density, $\rho=\rho_1+\rho_2$ is the
total mass density, ${\bf V}={\bf v}-{\bf U}$ is the peculiar velocity, and $p$ is the hydrostatic pressure. Furthermore, the third equality of Eq.\ (\ref{2.4}) defines the kinetic temperatures $T_i$ of each species, which measure  their mean kinetic energies. For inelastic collisions ($\al_{ij}\neq 1$), energy equipartition is in general broken and so $T_i \neq T$ \cite{G19}.

The collision operators conserve the particle number of each species and the
total momentum, but the total energy is not conserved. These conditions lead to the following identities:
\begin{equation}
\label{2.5}
\int\, d{\bf v} J_{ij}[{\bf v}|f_i,f_j]=0,
\end{equation}
\begin{equation}
\label{2.6}
\sum_{i=1}^2\sum_{j=1}^2\int \mathrm{d}{\bf v}m_{i}{\bf v} J_{ij}[{\bf v}|f_{i},f_{j}]=0,
\end{equation}
\begin{equation}
\label{2.7}
\sum_{i=1}^2\sum_{j=1}^2\int \mathrm{d}{\bf v}\frac{1}{2}m_{i}V^{2}J_{ij}
[{\bf v}|f_{i},f_{j}]=-\frac{d}{2}nT\zeta.
\end{equation}
Here, $\zeta$ is identified as the ``cooling rate'' due to inelastic
collisions among all species. At a kinetic level, it is also convenient to introduce
the ``cooling rates'' $\zeta_i$ for the partial temperatures $T_i$. They are defined as
\begin{equation}
\label{2.8}
\zeta_i=\sum_{j=1}^2 \zeta_{ij}=-\sum_{j=1}^2 \frac{1}{dn_iT_i}\int \mathrm{d}{\bf v}m_iV^{2}J_{ij}[{\bf v}|f_{i},f_{j}],
\end{equation}
where the second equality defines the quantities $\zeta_{ij}$. According to equations \eqref{2.7} and \eqref{2.8}, the total cooling rate $\zeta$ can be written in terms of the partial cooling rates $\zeta_i$ as
\begin{equation}
\label{2.9}
\zeta=T^{-1}\sum_{i=1}^2\; x_iT_i\zeta_i,
\end{equation}
where $x_i=n_i/n$ is the concentration or mole fraction of species $i$.

The macroscopic balance equations for the densities of mass, momentum and energy can be easily now derived from the constraints \eqref{2.5}--\eqref{2.7}. They are given by
\begin{equation}
D_{t}n_{i}+n_{i}\nabla \cdot {\bf U}+\frac{\nabla \cdot {\bf j}_{i}}{m_{i}}
=0\;,  \label{2.10}
\end{equation}
\begin{equation}
D_{t}{\bf U}+\rho ^{-1}\nabla \cdot \mathsf{P}=0\;,  \label{2.11}
\end{equation}
\begin{equation}
D_{t}T-\frac{T}{n}\sum_{i=1}^2\frac{\nabla \cdot \mathbf{j}_{i}}{m_{i}}+\frac{2}{dn}
\left( \nabla \cdot \mathbf{q}+\mathsf{P}:\nabla \mathbf{U}\right)
=-\zeta T\;. \label{2.12}
\end{equation}
In the above equations, $D_{t}=\partial _{t}+{\bf U}\cdot \nabla $ is the
material derivative,
\begin{equation}
\mathbf{j}_{i}=m_{i}\int \mathrm{d}{\bf v}\,{\bf V}\,f_{i}({\bf v})
\label{2.13}
\end{equation}
is the mass flux for species $i$ relative to the local flow,
\begin{equation}
\mathsf{P}=\sum_{i=1}^2\,\int \mathrm{d}{\bf v}\,m_{i}{\bf V}{\bf V}\,f_{i}({\bf  v})
\label{2.14}
\end{equation}
is the total pressure tensor, and
\begin{equation}
\mathbf{q}=\sum_{i=1}^2\,\int \mathrm{d}{\bf v}\,\frac{1}{2}m_{i}V^{2}{\bf V}
\,f_{i}({\bf v})
\label{2.15}
\end{equation}
is the total heat flux. It must be remarked that the balance equations (\ref{2.10})--(\ref{2.12}) apply regardless of the details of the model for inelastic collisions considered. However, the influence of the collision model appears through the dependence of the cooling rate and the hydrodynamic fluxes on the coefficients of restitution.

\subsection{Inelastic Maxwell models}

On the other hand, the hydrodynamic equations \eqref{2.10}--\eqref{2.12} do not constitute a closed set of differential equations for the hydrodynamic fields. To close them, one needs to solve the Boltzmann equation \eqref{2.1} to derive the corresponding constitutive equations for the fluxes and identify the explicit expressions of the transport coefficients. In the case of IHS, those expressions cannot be exactly determined and one has to resort to approximate solutions based on the use of Grad's moment method \cite{G49,JR85a,JR85b,G19} and/or the truncation of a Sonine polynomial expansion \cite{BP04,G19}. As mentioned in the Introduction section, a possible way of obtaining exact forms for the transport coefficients is to consider IMM where the collision rate is independent of the relative velocity of the colliding spheres. For this interaction model, the Boltzmann collision operators $J_{ij}^{\text{IMM}}[f_i,f_j]$ are \cite{BK03,G19}
\beq
J_{ij}^\text{IMM}[\mathbf{v}_1|f_i,f_j] =\frac{\nu_{\text{M},ij}}{n_jS_d}
\int \mathrm{d}{\bf v}_{2}\int \mathrm{d}\widehat{\boldsymbol{\sigma }}\Big[ \alpha_{ij}^{-1}f_{i}({\bf v}_{1}'')
f_{j}({\bf v}_{2}'')-f_{i}({\bf v}_{1})f_{j}({\bf v}_{2})\Big].
\label{2.16}
\eeq
Here, $\nu_{\text{M},ij}$ is an effective velocity-independent collision
frequency  for collisions  of type $i$-$j$ and  $S_d=2\pi^{d/2}/\Gamma(d/2)$ is the total solid angle in $d$ dimensions. In
Eq.\ \eqref{2.16}, in a binary collision the relationship between the pre-collisional velocities $(\mathbf{v}_1^{\prime\prime},\mathbf{v}_2^{\prime\prime})$ and the post-collisional velocities $(\mathbf{v}_1,\mathbf{v}_2)$ is
\begin{equation}
\label{2.17}
{\bf v}_{1}^{\prime\prime}={\bf v}_{1}-\mu_{ji}\left( 1+\alpha_{ij}^{-1}
\right)(\widehat{\boldsymbol{\sigma}}\cdot {\bf g}_{12})\widehat{\boldsymbol
{\sigma}}, \quad \quad {\bf v}_{2}^{\prime\prime}={\bf v}_{2}+\mu_{ij}\left(
1+\alpha_{ij}^{-1}\right) (\widehat{\boldsymbol{\sigma}}\cdot {\bf
g}_{12})\widehat{\boldsymbol{\sigma}}\;,
\eeq
where ${\bf g}_{12}={\bf v}_1-{\bf v}_2$ is the relative velocity of the colliding pair,
$\widehat{\boldsymbol{\sigma}}$ is a unit vector directed along the centers of the two colliding
spheres, and $\mu_{ij}=m_i/(m_i+m_j)$.
As for IHS \cite{BP04}, the scattering rules \eqref{2.17} yield $(\widehat{\boldsymbol{\sigma}}\cdot \mathbf{g}_{12}^{\prime\prime})=-\al_{ij}^{-1}(\widehat{\boldsymbol{\sigma}}\cdot \mathbf{g}_{12})$ where $\mathbf{g}_{12}^{\prime\prime}={\bf v}_1^{\prime\prime}-{\bf v}_2^{\prime\prime}$. The collision frequencies $\nu_{\text{M},ij}$ appearing in \eqref{2.16} can be seen as free parameters in the model. As usual, its dependence on the restitution coefficients $\alpha_{ij}$ and the parameters of the mixture can be chosen to optimize the agreement with  the results obtained from the Boltzmann equation  for IHS. Of course, the choice is not unique and may depend on the property of interest.

It is important to note that the inelastic Boltzmann equation applies only to very dilute gases. This means that the diameter of the hard spheres is much smaller than the mean free path between particles. As a consequence, all the distribution functions appearing in the Boltzmann collision operator \eqref{2.16} are evaluated at the same point $\mathbf{r}$ and at the same time $t$. Because of this, collisional transfer contributions to the pressure tensor $P_{ij}$ and the heat flux $\mathbf{q}$ are negligible in the low-density regime. Thus, only kinetic contributions to these fluxes are relevant. In addition, a crucial hypothesis in the derivation of the operator $J_{ij}[f_i,f_j]$ is the absence of correlations between the velocities of the two particles that are about to collide (molecular chaos hypothesis). This assumption is essential to obtaining the form of the Boltzmann operator \eqref{2.16} since the two-body distribution function $f_{2,ij}(\mathbf{r}, \mathbf{v}_1, \mathbf{r}, \mathbf{v}_2)$ factorizes into the product of the one-particle velocity distribution functions $f_{i}(\mathbf{r}, \mathbf{v}_1)$ and $f_{j}(\mathbf{r}, \mathbf{v}_1)$. More details on the derivation of the inelastic Boltzmann equation can be found in the textbooks \cite{BP04} and \cite{G19}.

As happens for elastic collisions \cite{CC70,GS03}, the main advantage of using IMM instead of IHS is
that a velocity moment of order $k$ of the Boltzmann collision operator $J_{ij}^\text{IMM}[f_i,f_j]$ only involves moments of order less than or equal to $k$ of the
distributions functions. This allows one to determine the Boltzmann collisional moments without the explicit
knowledge of the velocity distribution functions. In particular, the first and second collisional moments of  $J_{ij}^\text{IMM}[f_i,f_j]$ are \cite{G03bis}
\begin{equation}
\label{2.19}
\int \mathrm{d}{\bf v} m_i{\bf V}J_{ij}^\text{IMM}[f_i,f_j]=-\frac{\nu_{\text{M},ij}}{d\rho_j}\mu_{ji}(1+\alpha _{ij})
\left(\rho_j{\bf j}_i-\rho_i{\bf j}_j\right),
\end{equation}
\begin{eqnarray}
\label{2.20}
\int \mathrm{d}{\bf v} m_i V_k V_\ell J_{ij}^\text{IMM}[f_i,f_j]&=& -\frac{\nu_{\text{M},ij}}{d\rho_j}\mu_{ji}(1+\alpha _{ij})
\Bigg\{2\rho_j P_{i,k \ell}-\left(j_{i,k}j_{j,\ell}+j_{j,k}j_{i,\ell}\right)
\nonumber\\
& &-\frac{2}{d+2}\mu_{ji}(1+\alpha_{ij})\Bigg[\rho_jP_{i,k \ell}+\rho_iP_{j,k \ell}-
\left(j_{i,k}j_{j,\ell}+j_{j,k}j_{i,\ell}\right) \nonumber\\
& &+ \Bigg(\frac{d}{2}\left(\rho_i p_j+\rho_jp_i\right)-{\bf j}_i\cdot {\bf j}_j\Bigg)\delta_{k \ell}\Bigg]
\Bigg\},
\end{eqnarray}
where $p_i=n_i T_i=\text{Tr} (\mathsf{P}_i)/d$. The quantities $\zeta_{ij}$ defined by Eq.\ \eqref{2.8} can be exactly obtained for IMM from Eq.\ \eqref{2.20} as
\beq
\label{2.21}
\zeta_{ij}=\frac{2\nu_{\text{M},ij}}{d}\mu_{ji}(1+\alpha _{ij})\Bigg[1-\frac{\mu_{ji}}{2}(1+\alpha _{ij})\frac{\theta_i+\theta_j}{\theta_j}+\frac{\mu_{ji}(1+\alpha _{ij})-1}{d\rho_j p_i}{\bf j}_i\cdot {\bf j}_j\Bigg],
\eeq
where
\beq
\label{2.22}
\theta_i=\frac{m_i T}{\overline{m} T_i}, \quad \overline{m}=\frac{m_1+m_2}{2}.
\eeq

To optimize the agreement with the IHS results we adjust the collision frequencies $\nu_{\text{M},ij}$ to obtain the same expression of $\zeta_{ij}$ as the one found for IHS in the HCS \cite{G19}. However, given that the cooling rates are not exactly known for IHS, a good estimate for them can be achieved by considering the local equilibrium approximation for the velocity distribution functions $f_i$, i.e.,
\begin{equation}
\label{2.23}
f_i({\bf V})\to f_{i,\text{M}}({\bf V})=n_i\left(\frac{m_i}{2\pi T_i}\right)^{d/2}\exp\left(-\frac{m_iV^2}{2T_i}\right).
\end{equation}
In this approximation, one has \cite{G19}
\begin{equation}
\label{2.24}
\zeta_{ij}^{\text{IHS}}\to \frac{4\pi^{(d-1)/2}}{d\Gamma\left(\frac{d}{2}\right)}
n_j\mu_{ji}\sigma_{ij}^{d-1}\left(\frac{\theta_i+\theta_j}{\theta_i\theta_j}\right)^{1/2}
(1+\alpha_{ij})\left[1-\frac{\mu_{ji}}{2}(1+\alpha_{ij})
\frac{\theta_i+\theta_j}{\theta_j}\right]\upsilon_\text{th},
\end{equation}
where $\upsilon_\text{th}=\sqrt{2T/\overline{m}}$ is a thermal velocity defined in terms of the temperature $T(t)$ of the mixture. Thus, according to equations (\ref{2.21}) and \eqref{2.24}, to get $\zeta_{ij}=\zeta_{ij}^{\text{IHS}}$ one has to chose the collision frequencies $\nu_{\text{M},ij}$ as
\begin{equation}
\label{2.25}
\nu_{\text{M},ij}=
\frac{2\pi^{(d-1)/2}}{\Gamma\left(\frac{d}{2}\right)}n_j \sigma_{ij}^{d-1}\left(\frac{2{T}_i}{m_i}+
\frac{2{T}_j}{m_j}\right)^{1/2}.
\end{equation}
Upon deriving Eq.\ \eqref{2.25} use has been made of the fact that the mass flux ${\bf j}_i$ vanishes in the HCS. In the remainder of this paper, we will take the choice (\ref{2.25}) for $\nu_{\text{M},ij}$.

\subsection{Kinetic model for granular mixtures}

Another different way of overcoming the mathematical intricacies of the Boltzmann collision operator for IHS is to consider a kinetic model. Here, as said in section \ref{sec1}, we consider the VGS kinetic model \cite{VGS07} for granular mixtures. To the best of our knowledge, this is the only kinetic model has been proposed so far in the granular literature for
 this sort of systems.
The model is based on the equivalence between a system of \emph{elastic} hard spheres subject to a nonconservative force proportional to the particle velocity with a gas of IHS \cite{SA05}. This (approximate) mapping between a molecular hard sphere gas in the presence of a drag force with IHS allows to extend any kinetic model of molecular mixtures proposed in the literature to inelastic multicomponent gases. Here, the relaxation term appearing in the model reported in \cite{VGS07} has been chosen to be the well-known Gross-Krook (GK) kinetic model \cite{GK56} proposed many years ago for studying transport properties of multicomponent molecular gases. In this sense, the kinetic model employed in this paper can be seen as a direct extension of the GK model \cite{GK56} to granular mixtures.

In the tracer limit, within the VGS kinetic model for granular mixtures \cite{VGS07}, the Boltzmann collision operators $J_{22}[f_2,f_2]$ and $J_{12}[f_1,f_2]$ are defined respectively, as
\beq
\label{5.1}
J_{22}[f_2,f_2]\to K_{22}[f_2,f_2]=-\frac{1+\al_{22}}{2}\nu_{22}\left(f_2-f_{22}\right)+\frac{\epsilon_{22}}{2}\frac{\partial}{\partial \mathbf{V}}\cdot \Big[(\mathbf{v}-\mathbf{U}_2)\Big]f_2,
\eeq
\beq
\label{5.2}
J_{12}[f_1,f_2]\to K_{12}[f_1,f_2]=-\frac{1+\al_{12}}{2}\nu_{12}\left(f_1-f_{12}\right)+\frac{\epsilon_{12}}{2}\frac{\partial}{\partial \mathbf{V}}\cdot \Big[(\mathbf{v}-\mathbf{U}_1)\Big]f_1,
\eeq
where
\beq
\label{5.3}
\nu_{ij}=\frac{4\pi^{(d-1)/2}}{d\Gamma\left(\frac{d}{2}\right)}n_j\sigma_{ij}^{d-1}\left(\frac{2\widetilde{T}_i}{m_i}+
\frac{2\widetilde{T}_j}{m_j}\right)^{1/2}
\eeq
is an effective collision frequency for IHS, $\widetilde{T}_i=T_i-(m_i/d)(\mathbf{U}_i-\mathbf{U})^2$,
\beq
\label{5.4}
\epsilon_{ij}=\frac{1}{2}\nu_{ij}\mu_{ji}^2\left[1+\frac{m_i\widetilde{T}_j}{m_j
	\widetilde{T}_i}+\frac{3}{2d}\frac{m_i}{\widetilde{T}_i}\left({\bf
	U}_i-{\bf U}_j\right)^2\right](1-\alpha_{ij}^2),
\eeq
and
\beq
\label{5.5}
f_{ij}(\mathbf{v})=n_i\left(\frac{m_i}{2\pi
	T_{ij}}\right)^{d/2} \exp\left[-\frac{m_i}{2T_{ij}}\left(\mathbf{v}-\mathbf{U}_{ij}\right)^2\right].
\eeq
In Eq.\ \eqref{5.5}, we have introduced the quantities
\beq
\label{5.6}
\mathbf{U}_{ij}=\mu_{ij}\mathbf{U}_i+\mu_{ji}\mathbf{U}_j, \quad
T_{ij}=\widetilde{T}_i+2\mu_{ij}\mu_{ji}\Bigg\{\widetilde{T}_j-\widetilde{T}_i+\frac{(\mathbf{U}_i-\mathbf{U}_j)^2}{2d}
\left[m_j+
\frac{\widetilde{T}_j-\widetilde{T}_i}{\widetilde{T}_i/m_i+\widetilde{T}_j/m_j}\right]\Bigg\}.
\eeq
According to equations\ \eqref{2.25} and \eqref{5.3}, note that $\nu_{\text{M},ij}\neq \nu_{ij}$.


\section{Rheology of a sheared granular mixture in the tracer limit. IMM}
\label{sec3}

We consider the tracer limit ($x_1\to 0$) in a granular binary mixture. In this situation, since the concentration of the tracer species is negligibly small, the state of the excess granular gas is not perturbed by the presence of the tracer particles. Thus, the distribution function $f_2$ of the excess granular gas obeys the nonlinear closed Boltzmann equation:
\begin{equation}
\label{2.26}
\partial_t f_2+{\bf v}\cdot \nabla f_{2}=J_{22}^{\text{IMM}}\left[f_{2},f_{2}\right] \;.
\end{equation}
Additionally, the collisions between tracer particles themselves can be also neglected in the kinetic equation of the distribution $f_1$:
\begin{equation}
\label{2.27}
\partial_t f_1+{\bf v}\cdot \nabla f_{1}=J_{12}^{\text{IMM}}\left[f_{1},f_{2}\right] \;.
\end{equation}

We assume that the system (granular gas plus tracers) is under USF. At a macroscopic level, the USF state is characterized by constant densities $n_2\simeq n$ and $n_1$, a uniform granular temperature, and a linear velocity profile
\beq
\label{3.1}
U_{1,k}=U_{2,k}=U_k=a_{k\ell}r_\ell, \quad a_{k\ell}=a \delta_{kx}\delta_{\ell y},
\eeq
$a$ being the constant shear rate. This linear velocity profile assumes no boundary layer near the walls and is generated by the Lee-Edwards boundary conditions \cite{LE72}, which are simply periodic boundary conditions in the local Lagrangian frame moving with the mean flow velocity $\mathbf{U}$. For elastic collisions, the temperature grows in time due to the viscous heating term ($-aP_{xy}>0$) and hence a
steady state is not possible unless an external (artificial) force is introduced \cite{EM90}. However, in the case of inelastic collisions, the temperature changes in time due to the competition between two (opposite) mechanisms: on
the one hand, viscous (shear) heating ($-aP_{xy}>0$) and, on the other hand, energy dissipation in collisions ($-\zeta T<0$). A steady state
is achieved when both mechanisms cancel each other and the fluid autonomously seeks the temperature at which the
above balance occurs. Under stationary conditions, the balance equation (\ref{2.12}) becomes
\begin{equation}
\label{3.2}
aP_{2,xy}=-\frac{d}{2}\zeta_2 p_2,
\end{equation}
where we have taken into account that in the tracer limit, $P_{xy}\simeq P_{2,xy}$, $\zeta\simeq \zeta_2$ and $p\simeq p_2=n_2 T_2$. According to equation \eqref{3.2}, it is quite apparent the intrinsic connection
between the shear field and collisional dissipation in the system. Thus, the steady shear flow state characterized by Eq.\ \eqref{3.2} is inherently a non-Newtonian state since the collisional cooling (which is fixed by the mechanical properties of the particles) sets the strength of the (reduced) velocity gradient in the steady state. This means that for given values of the shear rate $a$ and the coefficient of restitution $\alpha_{22}$, the steady state relation \eqref{3.2} gives the (reduced) shear rate $a^*=a/\nu_{\mathrm{M},22}(T)$ as a {\em unique} function of the coefficient of restitution $\alpha_{22}$.

At a microscopic level, all the space dependence of the distribution functions $f_2(\mathbf{r}, \mathbf{v}; t)$ and $f_1(\mathbf{r}, \mathbf{v}; t)$ only occurs through their dependence on the peculiar velocity $\mathbf{V}=\mathbf{v}-\mathbf{U}$ \cite{DSBR86}. Thus, the USF is defined as that which is spatially \emph{homogeneous} when one refers the velocities of the particles to the local Lagrangian frame moving with the linear velocity field $U_k=a_{k\ell}r_\ell$. In this frame, the distributions $f_2$ and $f_1$ adopt the forms
\beq
\label{3.2.1}
f_2(\mathbf{r}, \mathbf{v}; t)=f_2(\mathbf{V};t ), \quad f_1(\mathbf{r}, \mathbf{v}; t)=f_1(\mathbf{V}; t).
\eeq
The fact that the USF becomes homogeneous in the above Lagrangian frame implies that it does not necessarily require the application of boundary conditions to be generated in computer simulations. However, as said before, the usual boundary conditions employed to generate the USF in simulations are the well-known Lees--Edwards periodic boundary conditions \cite{LE72}.

In the steady state, the corresponding set of Boltzmann equations \eqref{2.26} and \eqref{2.27} in the above Lagrangian frame become
\beq
\label{3.3}
-a V_y\frac{\partial f_2}{\partial V_x}=J_{22}^{\text{IMM}}\left[f_{2},f_{2}\right], \quad
-a V_y\frac{\partial f_1}{\partial V_x}=J_{12}^{\text{IMM}}\left[f_{1},f_{2}\right].
\eeq
Since the mass and heat fluxes vanish in the steady USF state, the relevant transport properties of the system are related to the pressure tensors $P_{2,k \ell}$ and $P_{1,k \ell}$ defined as
\beq
\label{3.4}
P_{i,k \ell}=\int \mathrm{d}\mathbf{V}\; m_i V_k V_\ell f_i(\mathbf{V}).
\eeq

\subsection{Rheology for the excess granular gas}

Explicit expressions for the (reduced) nonzero elements of $\mathsf{P}_2^*=\mathsf{P}_2/p_2$ and
$\mathsf{P}_1^*=\mathsf{P}_1/(x_1p_2)$ can be easily obtained from equations\ \eqref{3.3} when one takes into account Eq.\ \eqref{2.20} with $\mathbf{j}_i=\mathbf{0}$. In terms of the dimensionless quantities
\beq
\label{3.4.1}
\nu_\eta^*=\frac{(d+1-\al_{22})(1+\al_{22})}{d(d+2)}
\eeq
and
\beq
\zeta_2^*=\frac{\zeta_2}{\nu_{\text{M},22}}=\frac{1-\al_{22}^2}{2d},
\eeq
the nonzero elements of $\mathsf{P}_2^*$ are \cite{G03}
\beq
\label{3.5}
P_{2,yy}^*=P_{2,zz}^*=1-\frac{\zeta_2^*}{\nu_\eta^*}=\frac{d}{2}\frac{1+\al_{22}}{d+1-\al_{22}}, \quad P_{2,xx}^*=d-(d-1)P_{2,yy}^*=\frac{d}{2}\frac{d+3-(d+1)\al_{22}}{d+1-\al_{22}},
\eeq
\beq
\label{3.6}
P_{2,xy}^*=-\frac{P_{2,yy}^*}{\nu_\eta^*}a^*=-\frac{d^2(d+2)}{2(d+1-\al_{22})^2}a^*,
\eeq
\beq
\label{3.6.1}
a^*=\frac{a}{\nu_{\text{M},22}}=\sqrt{\frac{d}{2}\frac{\nu_\eta^*\zeta_2^*}{P_{2,yy}^*}}=
\frac{d+1-\al_{22}}{d}\sqrt{\frac{1-\al_{22}^2}{2(d+2)}}.
\eeq

As mentioned before, the expression \eqref{3.6.1} clearly indicates the intrinsic connection between the (reduced) shear rate and collisional dissipation in \textit{steady} USF. The parameter $a^*$ is the relevant nonequilibrium parameter of the problem because it measures the departure of the system from equilibrium. For elastic collisions ($\al_{22}=1$), $a^*=0$ and the equilibrium results of a molecular gas are recovered; that is, $P_{2,k\ell}^*=\delta_{k\ell}$. As the coefficient of restitution $\al_{22}$ decreases (collision dissipation increases), $a^*$ increases and so, the granular gas departs from equilibrium. According to equation \eqref{3.6.1}, since $0\leq \al_{22}\leq 1$, then the range of reduced shear rates $a^*$ is defined in the interval
$0\leq a^{*2} \leq (d+1-\al_{22})^2(1-\al_{22}^2)/(2d^2(d+2))$. In particular, for disks ($d=2$), $0\leq a^{*}\leq 0.53$ while $0\leq a^{*}\leq 0.42$ for spheres ($d=3$).

\subsection{Rheology for the tracer particles}

In the case of $\mathsf{P}_1^*$, its nonzero elements are given by \cite{G03bis}
\beq
\label{3.7}
P_{1,yy}^*=P_{1,zz}^*=-\frac{Y+X P_{2,yy}^*}{X_0}, \quad P_{1,xy}^*=\frac{a^*P_{1,yy}^*-XP_{2,xy}^*}{X_0},
\quad P_{1,xx}^*=d\gamma-(d-1)P_{1,yy}^*,
\eeq
where $\gamma=T_1/T_2$ is the temperature ratio and
\beq
\label{3.8}
Y=\frac{1}{d+2}\mu_{12}\mu_{21}\left(\frac{1+\theta}{\theta}\right)(1+\al_{12})^2
\nu_{\text{M},12}^*,
\eeq
\beq \label{3.8bis}
X_0=-\frac{2}{d(d+2)}\mu_{21}(1+\al_{12})\left[d+2-\mu_{21}(1+\al_{12})\right] \nu_{\text{M},12}^*,
\eeq
\beq
\label{3.9}
X=\frac{2}{d(d+2)}\mu_{12}\mu_{21}(1+\al_{12})^2 \nu_{\text{M},12}^*.
\eeq
In equations\ \eqref{3.8} and \eqref{3.9}, $\theta=\theta_1/\theta_2=\mu/\gamma$, $\mu=m_1/m_2$ is the mass ratio and
\beq
\label{3.10}
\nu_{\text{M},12}^*=\frac{\nu_{\text{M},12}}{\nu_{\text{M},22}}=\left(\frac{\sigma_{12}}{\sigma_2}\right)^{d-1}
\left(\frac{1+\theta}{2\theta}\right)^{1/2}.
\eeq
The temperature ratio is obtained by numerically solving the equation
\beq
\label{3.11}
\frac{P_{1,xy}^*}{P_{2,xy}^*}=\frac{\gamma \zeta_1^*}{\zeta_2^*},
\eeq
where
\beq
\label{3.12}
\zeta_1^*=\frac{2\nu_{\text{M},12}^*}{d}\mu_{21}(1+\al_{12})\Big[1-\frac{\mu_{21}}{2}(1+\al_{12})(1+\theta)\Big].
\eeq
Appendix \ref{appA} shows that the steady state solutions \eqref{3.5}--\eqref{3.7} are indeed linearly stable solutions.

\section{Tracer diffusion under USF: Results for IMM}
\label{sec4}

We assume that the USF state characterized by the rheological properties \eqref{3.5}--\eqref{3.7} is slightly perturbed by weak spatial gradients. Under these conditions, one expects that the mass flux of the intruder $\mathbf{j}_1$ has nonzero contributions due to the existence of the gradients $\nabla x_1$, $\nabla p$ and $\nabla T$. Additionally, the anisotropy induced by the strong shear flow in the velocity space gives rise to the presence of tensorial quantities ($D_{k\ell}$, $D_{p,k\ell}$ and $D_{T,k\ell}$) to describe the mass transport instead of the conventional scalar coefficients ($D$, $D_p$ and $D_T$) when the granular gas is in the HCS \cite{GD02}. The determination of the above tensors is the main goal of the present work.

As already made in a previous paper \cite{G07}, we have to start from the Boltzmann equation \eqref{2.27} with a general space and time dependence. Thus, we denote by $U_{s,k}=a_{k\ell}r_\ell$ the mean flow velocity of the {\em undisturbed} USF state. However, when we disturb the USF state, the true mean flow velocity is $U_k=U_{s,k}+\delta U_k$, where $\delta U_k$ is a small perturbation to $U_{s,k}$. Additionally, the true peculiar velocity is $\mathbf{c}=\mathbf{v}-\mathbf{U}=\mathbf{V}-\delta \mathbf{U}$, where now $\mathbf{V}=\mathbf{v}-\mathbf{U}_s$.

In the frame moving with the (undisturbed) mean velocity $\mathbf{U}_s$, equation \eqref{2.27} becomes
\begin{equation}
\label{4.1}
\frac{\partial}{\partial t}f_1-aV_y\frac{\partial}{\partial V_x}f_1+\left({\bf V}+{\bf U}_s\right)
\cdot \nabla f_1=J_{12}^\text{IMM}[f_1,f_2],
\end{equation}
where the derivative $\nabla f_1$ is taken at constant ${\bf V}$. The macroscopic balance equations associated with this disturbed USF state follows from the general equations (\ref{2.10}), (\ref{2.11}), and
(\ref{2.12}) when one takes into account that $\mathbf{U}=\mathbf{U}_s+\delta \mathbf{U}$. The result is
\begin{equation}
\label{4.2}
\partial_tn_1+\mathbf{U}_s\cdot \nabla n_1=-\nabla \cdot (n_1\delta \mathbf{U})-
\frac{\nabla \cdot {\bf j}_1}{m_1},
\end{equation}
\begin{equation}
\label{4.3}
\partial_t\delta U_k+a_{k\ell}\delta U_\ell+(U_{s,\ell}+\delta U_\ell) \nabla_\ell \delta U_k=-
\rho^{-1}\nabla_j P_{2,k\ell},
\end{equation}
\begin{equation}
\label{4.4} \frac{d}{2}n\partial_tT+\frac{d}{2}n(\mathbf{U}_s+\delta \mathbf{U})\cdot \nabla T+aP_{xy}+\nabla \cdot
{\bf q}+{\sf P}:\nabla \delta \mathbf{U}=-\frac{d}{2}p\zeta.
\end{equation}
Here, we recall that in the tracer limit $\rho\simeq \rho_2=m_2 n_2$, $n\simeq n_2$, $T\simeq T_2$, $\zeta\simeq \zeta_2$, and $\mathsf{P}\simeq \mathsf{P}_2$. In equations \eqref{4.2}--\eqref{4.4}, the expressions of the cooling rate $\zeta$, the mass flux ${\bf j}_1$, the pressure tensor ${\sf P}$, and the heat flux ${\bf q}$ are defined by equations (\ref{2.7}), (\ref{2.13}), (\ref{2.14}), and (\ref{2.15}), respectively, with the replacement of ${\bf V}$ by the peculiar velocity $\mathbf{c}$ in the disturbed USF state .

\subsection{Generalized Chapman--Enskog expansion}

Since the deviations from the USF are assumed to be small, our goal is to solve the Boltzmann--Lorentz kinetic equation \eqref{4.1} up to first order in the spatial gradients of the hydrodynamic fields
\beq
\label{4.5}
A({\bf r};t)\equiv \{x_1({\bf r},t), p({\bf r}, t), T({\bf r}, t), \delta {\bf U}({\bf r},t)\}.
\eeq
In equation \eqref{4.5}, as in previous works on granular mixtures \cite{GD02}, we represent the mass flux $\mathbf{j}_1$ in terms of the spatial gradients of the fields $x_1$, $p=n T$, and $T$. However, in contrast to previous studies \cite{GD02,SGNT06} where the granular gas is in the HCS, the system is strongly sheared and hence the conventional Chapman--Enskog method \cite{CC70} cannot be applied. Thus, as in the paper \cite{G07}, we look for a solution to equation (\ref{4.1}) by using a generalized Chapman--Enskog--like expansion where the velocity
distribution function is expanded about a {\em local} shear flow reference state in terms of the small spatial
gradients of the hydrodynamic fields relative to those of USF. This is the main new ingredient of the expansion.
This type of generalized Chapman--Enskog expansion has been considered in the case of elastic gases to get the set of
shear-rate dependent transport coefficients \cite{LD97} in a thermostatted shear flow problem and it has
also been considered for monocomponent \cite{L06,G06a} and multicomponent \cite{G07,GT15} granular gases under USF.

As in the conventional Chapman--Enskog method \cite{CC70}, we look for a \textit{normal} or hydrodynamic solution. This means that the spatial dependence of the distribution $f_1$ occurs only through a functional dependence of the hydrodynamic fields $A({\bf r},t)$. Thus, $f_1(\mathbf{r}, \mathbf{v}; t)=f_1(A(\mathbf{r};t), \mathbf{V})$. We can make this functional dependence local by assuming that the strength of the hydrodynamic fields is small, in which case $f_1$ can be expanded in powers of the fields as
\beq
\label{n1}
f_1(A(\mathbf{r};t), \mathbf{V})=f_1^{(0)}(A(\mathbf{r};t),\mathbf{V})+f_1^{(1)}(A(\mathbf{r};t),\mathbf{V})+\ldots.
\eeq
The successive approximations $f_1^{(k)}$ are of order $k$ in the gradients of $x_1$, $p$, $T$, and $\delta \mathbf{U}$, but they retain all the hydrodynamic orders in the shear rate.
In steady USF, this is equivalent to arbitrary values of the coefficient of restitution $\al_{22}$ due to the relation \eqref{3.6.1}. In this paper, only the first-order approximation will be considered. Substituting the expansion \eqref{n1} into the definitions \eqref{2.7} and \eqref{2.13}--\eqref{2.15} yields the corresponding  expansions for the cooling rate and the fluxes. Similarly to the conventional Chapman--Enskog method \cite{CC70}, the operator $\partial_t$ must be also expanded as
\begin{equation}
\label{n2}
\partial_t=\partial_t^{(0)}+\partial_t^{(1)}+\partial_t^{(2)}+\cdots,
\end{equation}
where the action of each operator $\partial_t^{(k)}$ is obtained from the hydrodynamic equations
(\ref{4.2})--(\ref{4.4}). These results provide the basis for generating the Chapman-Enskog solution to the inelastic Boltzmann--Lorentz equation (\ref{4.1}).

Before applying the Chapman--Enskog--like expansion to first order in spatial gradients, it is helpful to briefly discuss the convergence of this perturbation method.
To put the discussion in a proper context, it is important first to remark that, in spite of its long history and obvious importance, there are few exact results regarding the convergence of the conventional Chapman-Enskog expansion \cite{CC70} for molecular gases. To the best of our knowledge, the first study on this issue is due to Ikenberry and Truesdell \cite{IT56} for a gas of Maxwell molecules in USF. They obtained an expression of the stress tensor as a function of the shear rate that is analytic about the origin. However, for hard spheres, a solution of the BGK kinetic model of USF shows that the pressure tensor expansion in powers of the shear rate about the origin does not converge \cite{SBD86}. Conversely, a convergent expansion about the point at infinity suggests that the Chapman-Enskog expansion is asymptotic. Another interesting study is that of McLennan \cite{M65} for a general class of cutoff potentials. He proves the convergence of the expansion for the \textit{linearized} Boltzmann equation (namely, the convergence of the \textit{partial} sum of the Chapman--Enskog series made of the linear terms of the form $\partial_{\ell_1}\partial_{\ell_2}\ldots \partial_{\ell_k}A$, where $A$ denotes the hydrodynamic fields). Finally, a general discussion of Grad \cite{G63} indicates that the expansion is at least asymptotic.

Studies on the convergence of the Chapman--Enskog expansion for granular gases are much  scarcer than those for molecular gases. We are only aware of one study \cite{S08} that has analyzed this problem for sheared granular gases. Unlike the elastic case, starting from a simple kinetic model of IHS, it is shown that the expansion of the shear stress in powers of the shear rate is convergent. The radius of convergence increases with inelasticity. Studying the convergence of the generalized Chapman--Enskog expansion proposed in this paper is an interesting problem that goes beyond the scope of this work. As in the conventional Chapman--Enskog method \cite{CC70}, we expect the results derived here for the mass flux to be a well-behaved representation of the shear-rate dependence of the tracer diffusion coefficients to first order in spatial gradients (Navier--Stokes order). Qualitative agreement between kinetic theory results (based on the above Chapman--Enskog--like expansion) and molecular dynamics simulations for the self-diffusion tensor in dense granular gases supports the above statement.

\subsection{Zeroth-order approximation}

The kinetic equation for the zeroth-order distribution $f_1^{(0)}$ (reference distribution) can be obtained by substituting the expansions \eqref{n1}--\eqref{n2} into Eq.\ (\ref{4.1}):
\begin{equation}
\label{n3}
\partial_t^{(0)}f_1^{(0)}-aV_y\frac{\partial}{\partial V_x}f_1^{(0)}=J_{12}^{\text{IMM}}[{\bf
V}|f_1^{(0)},f^{(0}].
\end{equation}
To lowest order in the expansion the conservation laws give
\beq
\label{4.9}
\partial_t^{(0)}x_1=0, \quad  \partial_t^{(0)}\delta U_k=-a_{k \ell}\delta U_\ell,
\eeq
\beq
\label{4.10}
\partial_t^{(0)} \ln T=\partial_t^{(0)} \ln p=-\frac{2}{d p}a P_{xy}^{(0)}-\zeta.
\eeq
As shown in Refs.\ \cite{L06,G06a}, for given values of $a$ and $\alpha_{22}$, the steady-state condition (\ref{3.2})
establishes a mapping between the pressure $p$ and temperature $T$, such that every pressure corresponds to one and
only one temperature. However, in the {\em local} USF state (reference state) the pressure and temperature are specified separately. Hence, the collisional cooling term $\zeta p$ cannot be compensated for by the the viscous heating term $a|P_{xy}^{(0)}|$. This implies that the
zeroth-order distributions for the gas $f^{(0)}$ and the tracer $f_1^{(0)}$ depend on time through their dependence on the pressure and temperature. Consequently, the (reduced) shear rate $a^*=a/\nu_{\text{M},22}(p,T)$ depends on space and time through its dependence on the collision frequency $\nu_{\text{M},22}(p,T)$.

Equation \eqref{n3} can be rewritten in a more convenient form when one takes into account that the time dependence of $f_1^{(0)}$ is only through $x_1$, $p$, $T$, and $\delta \mathbf{U}$. Using equations \eqref{4.9}--\eqref{4.10}, one gets the kinetic equation
\begin{equation}
\label{n4}
-\left(\frac{2}{dp}a P_{xy}^{(0)}+\zeta\right)\left(p\frac{\partial}{\partial
p}+T\frac{\partial}{\partial T}\right)f_1^{(0)} -ac_y\frac{\partial}{\partial c_x}f_0^{(0)}=J_{12}^{\text{IMM}}[{\bf V}|f_1^{(0)},f^{(0}].
\end{equation}
Equation \eqref{n4} has the same form as the corresponding Boltzmann equation for the USF. On the other hand, in equation \eqref{n4} $f_1^{(0)}(\mathbf{r}, \mathbf{v}; t)$ is a \textit{local} USF distribution function. Thus, the occurrence of this local USF distribution as a reference state of the generalized Chapman--Enskog method is not an assumption and/or choice but rather a consequence of the solution to the Boltzmann--Lorentz equation to zeroth-order in the gradients. In the steady state ($\frac{2}{dp}a P_{xy}^{(0)}+\zeta=0$), the first term on the left hand side of equation \eqref{n4} vanishes and hence, it is easy to see that the expressions of the zeroth-order pressure tensor $P_{1,xy}^{(0)}$ are given by equations \eqref{3.7}--\eqref{3.9}.

\subsection{First-order approximation}

Given that most of mathematical steps involved in the determination of the first-order
contribution $\mathbf{j}_1^{(1)}$ to the mass flux for IMM are similar to those made in the paper \cite{G07} for IHS, only some partial results on this calculation will be reported in this section. We refer the interested reader to the work \cite{G07} for more specific details. The first-order distribution $f_1^{(1)}$ obeys the kinetic equation
\begin{align}
\label{4.7.1}
\left(\partial_t^{(0)}-aV_y\frac{\partial}{\partial V_x}\right)f_1^{(1)}-J_{12}^\text{IMM}[f_1^{(1)},f_2^{(0)}]=
{\bf A}_1\cdot \nabla x_1+{\bf B}_1\cdot \nabla p+{\bf C}_1\cdot \nabla T  +\,\mathsf{D}_1:\nabla \delta {\bf U} +J_{12}^\text{IMM}[f_1^{(0)},f_2^{(1)}].
\end{align}
In equation \eqref{4.7.1}, the quantities $\mathbf{A}_1(\mathbf{c})$, $\mathbf{B}_1(\mathbf{c})$, $\mathbf{C}_1(\mathbf{c})$, and $\mathsf{D}_1(\mathbf{c})$  are defined by equations\ (C.9)--(C.12), respectively, of \cite{G07} while the first-order distribution of the excess granular gas $f_2^{(1)}(\mathbf{c})$ has the form \cite{L06,G06a}
\beq
\label{4.8.1}
f_2^{(1)}(\mathbf{c})=
{\boldsymbol {\mathcal B}}_{2}(\mathbf{c})\cdot \nabla p+{\boldsymbol {\mathcal C}}_{2}(\mathbf{c})\cdot \nabla T+{\sf {\mathcal D}}_{2}(\mathbf{c}):\nabla\delta {\bf U}.
\eeq
According to the right hand side of Eq.\ \eqref{4.7.1}, the solution to this kinetic equation is
\begin{equation}
f_1^{(1)}(\mathbf{c})={\boldsymbol {\mathcal A}}_{1}(\mathbf{c})\cdot \nabla x_1+
{\boldsymbol {\mathcal B}}_{1}(\mathbf{c})\cdot \nabla p+{\boldsymbol {\mathcal C}}_{1}(\mathbf{c})\cdot \nabla T+{\sf {\mathcal D}}_{1}(\mathbf{c}):\nabla\delta {\bf U},
\label{4.8.2}
\end{equation}
where the coefficients ${\boldsymbol {\mathcal A}}_{1}$, ${\boldsymbol {\mathcal B}}_{1}$, ${\boldsymbol {\mathcal C}}_{1}$,
and ${\boldsymbol {\mathcal D}}_{1}$ are functions of the peculiar velocity and the
hydrodynamic fields $x_1$, $p$, and $T$.

To first-order, the mass flux $\mathbf{j}_1^{(1)}$ is defined as
\beq
\label{4.6}
\mathbf{j}_1^{(1)}=\int \mathrm{d}\mathbf{v}\; m_1 \mathbf{c} f_1^{(1)}(\mathbf{c}).
\eeq
Substitution of equation \eqref{4.8.2} into equation \eqref{4.6} gives the expression
\begin{equation}
\label{4.7}
j_{1,k}^{(1)}=-m_1 D_{k \ell}\nabla_\ell x_1-\frac{m_2}{T}D_{p,k \ell}\nabla_\ell p-\frac{\rho}{T}D_{T,k \ell}\nabla_\ell T,
\end{equation}
where $\nabla_\ell\equiv \partial/\partial r_\ell$ and
\begin{equation}
\label{4.8} D_{k \ell}=-\int \mathrm{d}{\bf c}\,c_k\;{\mathcal A}_{1,\ell}({\bf c}),
\end{equation}
\begin{equation}
  \quad D_{p,k \ell}=-\frac{Tm_1}{m_2}\int \mathrm{d}{\bf c}\,c_k\;{\mathcal B}_{1,\ell}({\bf c}),
\end{equation}
\begin{equation}
  D_{T,k \ell}=-\frac{Tm_1}{\rho}\int \mathrm{d}{\bf c}\,c_k\;{\mathcal C}_{1,\ell}({\bf c}).
\end{equation}
According to equation (\ref{4.7}), the mass flux of the tracer particles is
expressed in terms of a diffusion tensor $D_{k \ell}$, a pressure diffusion tensor $D_{p,k \ell}$, and a thermal diffusion tensor $D_{T,k \ell}$. In the time-dependent problem, the set of {\em generalized} transport coefficients $D_{k \ell}$, $D_{p,k \ell}$, and $D_{T,k \ell}$ are
nonlinear functions of the shear rate  and the parameters of the system (masses $m_1$ and $m_2$ and diameters $\sigma_1$ and $\sigma_2$, and the coefficients of restitution $\alpha_{22}$ and $\alpha_{12}$). Compared to the conventional diffusion problem in the Navier-Stokes domain \cite{GD02}, there are important differences. First, the diagonal elements of the diffusion tensors are generally different. Second, the off-diagonal elements (which are zero in the Navier--Stokes approximation) are different from zero. These features are both intrinsically related to the fact that the system is arbitrarily far from equilibrium. Interestingly, shear flow induces cross-effects in the diffusion of tracer particles. As we will show later, the elements $D_{xy}$, $D_{p,xy}$, and $D_{T,xy}$ are clearly different from zero for finite shear rates (which is equivalent to values of $\al_{22}$ different from 1 in the steady state). Thus, a gradient of concentration, pressure and/or temperature along the direction of the flow of the system ($x$-axis) creates a transport of tracers parallel to the gradient of the flow velocity ($y$-axis).

The corresponding integral equations for the unknowns ${\boldsymbol {\mathcal A}}_{1}$, ${\boldsymbol {\mathcal B}}_{1}$, and  ${\boldsymbol {\mathcal C}}_{1}$ defining the diffusion coefficients can be obtained by substituting equations\ \eqref{4.8.1} and \eqref{4.8.2} into equation \eqref{4.7.1} and identifying coefficients of independent gradients. To achieve these equations, one needs to take into account the action of the operator $\partial_t^{(0)}$ on the hydrodynamic fields given by equations \eqref{4.2}--\eqref{4.4}. The set of coupled linear integral equations for the unknowns can be obtained after considering the contributions coming from the action of the operator $\partial_t^{(0)}$ on $\nabla p$ and $\nabla T$:
\beq
\label{4.11}
\partial_t^{(0)} \nabla p=-\nabla \left(\frac{2}{d}a
P_{xy}^{(0)}+p\zeta\right)
=-\left(\frac{2a}{d}\frac{\partial P_{xy}^{(0)}}{\partial p}+ 2\zeta\right) \nabla
p-\left(\frac{2a}{d}\frac{\partial P_{xy}^{(0)}}{\partial T}-\frac{1}{2} \frac{p\zeta}{T}\right) \nabla T,
\eeq
\begin{align}
\label{4.12}
\partial_t^{(0)} \nabla T  =-\nabla \left(\frac{2T}{d p}a
P_{xy}^{(0)}+\zeta\right) =\left[\frac{2aT}{dp^2}\left(1-p\frac{\partial}{\partial p}\right)P_{xy}^{(0)}-\frac{T\zeta}{p}\right]
\nabla
p-\left[\frac{2a}{dp}\left(1+T\frac{\partial}{\partial T}\right)P_{xy}^{(0)}+\frac{1}{2}\zeta\right]
\nabla T.
\end{align}
Since  the distribution function  $f_1^{(1)}$ is qualified as a \emph{normal} or hydrodynamic solution, then it depends on time through its dependence on the hydrodynamic fields $A(\mathbf{r},t)$. As a consequence, for a given function $\mathcal{Z}(x_1, p, T, \delta \mathbf{U})$, one has the identity
\begin{eqnarray}
\label{4.13}
\partial_t^{(0)}\mathcal{Z} =\frac{\partial \mathcal{Z}}{\partial p}\partial_t^{(0)}
p+\frac{\partial \mathcal{Z}}{\partial T}\partial_t^{(0)} T+\frac{\partial \mathcal{Z}}{\partial \delta U_i}\partial_t^{(0)}
\delta U_i =-\left(\frac{2}{d p}a P_{xy}^{(0)}+\zeta\right)\left(p\frac{\partial}{\partial
p}+T\frac{\partial}{\partial T}\right) \mathcal{Z}+a_{k \ell}\delta U_\ell \frac{\partial \mathcal{Z}}{\partial c_k},
\end{eqnarray}
where in the last step we have taken into account that $\mathcal{Z}$ depends on $\delta {\bf U}$ through ${\bf c}={\bf
V}-\delta {\bf U}$.

\subsection{Steady state conditions}

As discussed in previous works \cite{G07,GT15}, the determination of the diffusion transport coefficients requires in general to numerically solve a set of differential equations. Thus, to get analytical forms for those coefficients, we restrict ourselves to linear deviations from the \emph{steady} USF state. In this case, since the contributions to the mass flux are already of first order in the deviations from this steady state, we have to compute $D_{ij}$, $D_{p,ij}$, and $D_{T,ij}$ to zero order in the deviations, namely, under steady state conditions. In this case, according to Eq.\eqref{4.10}, in the steady USF state $\partial_t^{(0)}p=\partial_t^{(0)}T=0$ which implies that $(2/dp)a P_{xy}^{(0)}+\zeta=0$. Therefore, taking into account the results \eqref{4.11}--\eqref{4.13}, the set of integral equations that the unknowns $\mathcal{X}\equiv \{ {\boldsymbol {\mathcal A}}_{1}$, ${\boldsymbol {\mathcal B}}_{1}, {\boldsymbol {\mathcal C}}_{1} \}$ obey in the steady state is
\begin{equation}
\label{4.14}
- a c_y\frac{\partial}{\partial c_x}{\boldsymbol {\mathcal A}}_{1}-J_{12}^\text{IMM}[{\boldsymbol {\mathcal
A}}_{1},f_2^{(0)}]={\bf A}_{1},
\end{equation}
\begin{align}
\label{4.15}
- a c_y\frac{\partial}{\partial c_x}{\boldsymbol {\mathcal B}}_{1}
- \left(\frac{2a}{d}\partial_p P_{2,xy}^{(0)}+2\zeta\right){\boldsymbol {\mathcal B}}_{1}
-J_{12}^\text{IMM}[{\boldsymbol {\mathcal B}}_{1},f_2^{(0)}]
=\mathbf{B}_1-\Bigg[\frac{2aT}{d p^2}\left(1-p\partial_p\right)P_{2,xy}^{(0)}-\frac{T\zeta}{p}\Bigg]
{\boldsymbol {\mathcal C}}_{1}+J_{12}^\text{IMM}[f_1^{(0)},{\boldsymbol {\mathcal B}}_{2}],
\end{align}
\begin{align}
\label{4.16}
- a c_y\frac{\partial}{\partial c_x}{\boldsymbol {\mathcal C}}_{1}
-\Bigg[\frac{2a}{d p}\left(1+T\partial_T\right)P_{2,xy}^{(0)}+\frac{1}{2}\zeta\Bigg]{\boldsymbol {\mathcal C}}_{1}
-J_{12}^\text{IMM}[{\boldsymbol {\mathcal C}}_{1},f_2^{(0)}] =\mathbf{C}_1+\left(\frac{2a}{d}\partial_T P_{2,xy}^{(0)}-\frac{p\zeta}{2T}\right){\boldsymbol {\mathcal B}}_{1}
+J_{12}^\text{IMM}[f_1^{(0)},{\boldsymbol {\mathcal C}}_{2}].
\end{align}
It must be recalled that in equations\ (\ref{4.14})--(\ref{4.16}) all the quantities are evaluated in the steady
state. In particular, the expressions of $\zeta$ and $P_{2,xy}^{(0)}$ are given by equations \eqref{3.4.1} and \eqref{3.6}, respectively. Henceforth, the calculations will be restricted to this particular condition.

As expected the structure of the integral equations (\ref{4.14})--(\ref{4.15}) is similar to the one derived for IHS \cite{G07}, except by the replacement of the Boltzmann--Lorentz collision operators $J_{12}^\text{IMM}[\mathcal{X},f_2^{(0)}]$ and $J_{12}^\text{IMM}[f_1^{(0)},\mathcal{X}]$ by their corresponding IHS counterparts. The main advantage of using IMM instead of IHS is that the collisional moments associated with the operator $J_{12}^\text{IMM}[\mathcal{X},\mathcal{Y}]$ can be \emph{exactly} computed. In particular, in the case of the mass flux, according to equation \eqref{2.19} one has the results
\beq
\label{4.17}
\int \mathrm{d}{\bf c} \; m_1 c_k
\left(
\begin{array}{c}
J_{12}^\text{IMM}[\mathcal{A}_{1,\ell},f_2^{(0)}]\\
J_{12}^\text{IMM}[\mathcal{B}_{1,\ell},f_2^{(0)}]\\
J_{12}^\text{IMM}[\mathcal{C}_{1,\ell},f_2^{(0)}]
\end{array}
\right)
=\Omega
\left(
\begin{array}{c}
m_1 D_{k \ell}\\
\frac{m_2}{T}D_{p,k \ell}\\
\frac{\rho}{T}D_{T,k \ell}
\end{array}
\right),
\eeq
\beq
\label{4.18}
\int \mathrm{d}{\bf c} \; m_1 c_k
\left(
\begin{array}{c}
J_{12}^\text{IMM}[f_1^{(0)},\mathcal{B}_{2,\ell}]\\
J_{12}^\text{IMM}[f_1^{(0)},\mathcal{C}_{2,\ell}]
\end{array}
\right)
=
\left(
\begin{array}{c}
0\\
0
\end{array}
\right),
\eeq
where we have taken into account that $\mathbf{j}_1^{(0)}=\mathbf{j}_2^{(1)}=\mathbf{0}$ and
\beq
\label{4.22}
\Omega=\frac{\nu_{\text{M},12}}{d}\mu_{21}(1+\al_{12}).
\eeq

Thus, multiplying both sides of equations\ \eqref{4.14}--\eqref{4.16} by $m_1 c_k$ and integrating over $\mathbf{c}$, one gets the set of algebraic equations:
\beq
\label{4.19}
\left(a_{k\mu}+\Omega \delta_{k \mu}\right)D_{\mu \ell}=\frac{n}{\rho_1}P_{1,k \ell}^{(0)},
\eeq
\begin{align}
\label{4.20}
\left(\frac{2a}{d}\partial_p P_{2,xy}^{(0)}+2\zeta\right)D_{p,k \ell}-\left(a_{k\mu}+\Omega \delta_{k \mu}\right)D_{p,\mu \ell}=
\frac{T}{m_2}\left(\frac{\rho_1}{\rho}\partial_p P_{2,k\ell}^{(0)}-\partial_p P_{1,k\ell}^{(0)}\right) +
\Bigg[\frac{2a}{d p}\left(1-p\partial_p\right)P_{2,xy}^{(0)}-\zeta\Bigg]D_{T,k\ell},
\end{align}
\begin{align}
\label{4.21}
\Bigg[\frac{2a}{d p}\left(1+T\partial_T\right)P_{2,xy}^{(0)}+\frac{1}{2}\zeta\Bigg]D_{T,k\ell}-\left(a_{k\mu}+\Omega \delta_{k \mu}\right)D_{T,\mu \ell}=\frac{T}{\rho}\left(\frac{\rho_1}{\rho}\partial_T P_{2,k\ell}^{(0)}-\partial_T P_{1,k\ell}^{(0)}\right)
-\left(\frac{2a}{d n}\partial_T P_{2,xy}^{(0)}-\frac{\zeta}{2}\right)D_{p,k\ell}.
\end{align}
In equations\ \eqref{4.20} and \eqref{4.21}, the derivatives of the pressure tensors with respect to the hydrostatic pressure $p$ and/or the temperature $T$ can be conveniently written in terms of the derivatives with respect to the (reduced) shear rate $a^*$ when one takes into account that $P_{2,k\ell}^{(0)}$ and $P_{1,k\ell}^{(0)}$ depend on $p$ and $T$ explicitly and also through their dependence on the reduced shear rate $a^*(p,T)\propto \sqrt{T}/p$. Thus,
\beq
\label{4.22.1}
\partial_p P_{2,k\ell}^{(0)}=\partial_p p P_{2,k\ell}^{*}(a^*)=\left(1-a^*\frac{\partial}{\partial a^*}\right)P_{2,k\ell}^{*}(a^*),
\eeq
\beq
\partial_T P_{2,k\ell}^{(0)}=\partial_T p P_{2,k\ell}^{*}(a^*)=\frac{p}{2T}a^*\frac{\partial}{\partial a^*}P_{2,k\ell}^{*}(a^*),
\eeq
\beq
\label{4.23}
\partial_p P_{1,k\ell}^{(0)}=\partial_p x_1 p P_{1,k\ell}^{*}(a^*)=x_1\left(1-a^*\frac{\partial}{\partial a^*}\right)P_{1,k\ell}^{*}(a^*),
\eeq
\beq
\partial_T P_{1,k\ell}^{(0)}=\partial_T x_1 p P_{1,k\ell}^{*}(a^*)=\frac{x_1 p}{2T}a^*\frac{\partial}{\partial a^*}P_{1,k\ell}^{*}(a^*).
\eeq
The derivatives of $P_{2,k\ell}^{*}$ and $P_{1,k\ell}^{*}$ with respect to $a^*$ in the steady state for IMM are obtained in the Appendix \ref{appB}.

For elastic collisions ($\al_{22}=\al_{12}=1$), equations\ \eqref{3.5}--\eqref{3.12} lead to $a^*=0$, $\zeta_2^*=0$, $P_{2,k\ell}^{*}=P_{1,k\ell}^{*}=\delta_{k\ell}$, $\gamma=1$, and
\beq
\label{4.25}
\Omega\equiv \Omega^{\text{el}}=\frac{4\pi^{(d-1)/2}}{d\Gamma\left(\frac{d}{2}\right)}n \sigma_{12}^{d-1}\sqrt{\frac{2Tm_2}{m_1(m_1+m_2)}}.
\eeq
In this limiting case, the solution to equations\ \eqref{4.19}--\eqref{4.21} is $D_{k\ell}=D^{\text{el}} \delta_{k\ell}$, $D_{p,k\ell}=D_p^{\text{el}} \delta_{k\ell}$, and $D_{T,k\ell}=D_T^{\text{el}} \delta_{k\ell}$, where
\beq
\label{4.26}
D^{\text{el}}=\frac{p}{m_1 \Omega^{\text{el}}}, \quad D_p^{\text{el}}=x_1\left(1-\frac{m_1}{m_2}\right)\frac{T}{m_2 \Omega^{\text{el}}}, \quad D_T^{\text{el}}=0.
\eeq
As expected, the expressions \eqref{4.26} agree with those obtained in the Navier--Stokes domain for a molecular mixture of Maxwell molecules in the tracer limit \cite{CC70}.

As expected from the results derived for IHS \cite{G07}, the coefficients $D_{k\ell}$ obey an autonomous set of algebraic equations whose solution can be written as
\beq
\label{4.24}
D_{k\ell}=\frac{p}{m_1 \Omega}\Bigg(\delta_{k\mu}-\frac{a_{k\mu}}{\Omega}
\Bigg)P_{1,\mu \ell}^*,
\eeq
where we recall that the tensor $a_{k\mu}=a\delta_{kx}\delta_{\mu y}$. On the other hand, since the coefficients $D_{p,k\ell}$ and $D_{T,k\ell}$ obey a set of algebraic \emph{coupled} equations, their forms are more intricate than that of the tracer diffusion tensor $D_{k\ell}$.

According to equation \eqref{4.24}, it is apparent that $D_{xx}\neq D_{yy}=D_{zz}$ while the only nonzero (negative) off-diagonal elements are $D_{xy}\neq D_{yx}$. This is a consequence of the symmetry of the USF problem. Additionally, for small shear rates, $P_{1,xy}^*\propto a^*$, so the off-diagonal elements are proportional to $a^*$ in this limiting situation. Thus, while the diagonal coefficients can be considered generalizations of the Navier--Stokes transport coefficients, the off-diagonal coefficients $D_{xy}$ and $D_{yx}$ can be considered generalizations of the Burnett
transport coefficients (i.e., coefficients relating the mass flux with terms of second order in the hydrodynamic gradients). A similar shear-rate dependence occurs for the elements of the tensors $D_{p,k\ell}$ and $D_{T,k\ell}$.

\subsection{Self-diffusion tensor}

An interesting limiting case corresponds to the self-diffusion problem, namely, when the tracer particles are mechanically equivalent to the particles of the granular gas (i.e., $\sigma_1=\sigma_2=\sigma$, $m_1=m_2=m$, and $\al_{22}=\al_{12}=\al$). This situation has been widely studied in computer simulations of dense granular gases \cite{C97,ZPTMS98,ALJR21}. In this limiting case, equations \ \eqref{4.20}--\eqref{4.21} lead to $D_{p,k\ell}=D_{T,k\ell}=0$ and the mass flux obeys a generalized Fick's law given by
\beq
\label{4.26.1}
j_{1,k}^{(1)}=-m D_{k\ell}^{\text{self}} \frac{\partial x_1}{\partial r_\ell},
\eeq
where
\beq
\label{4.26.2}
D_{k\ell}^{\text{self}}=\frac{p}{m \Omega}\Bigg(\delta_{k\mu}-\frac{a_{k\mu}}{\Omega}
\Bigg)P_{2,\mu \ell}^*,
\eeq
and $\Omega=(\nu_\text{M,22}/2d)(1+\al)$.

\section{Tracer diffusion under USF from a kinetic model of granular mixtures}
\label{sec5}

To complement the results derived for IMM we consider here the kinetic model (referred to as the VGS model) defined by equations\ \eqref{5.1} and \eqref{5.2}. Let us consider separately the results obtained in the USF for the excess granular gas and the tracer particles.

\subsection{Excess granular gas}

The rheological properties in the USF state can be easily determined from the first equation of \eqref{3.3} when one makes the replacement $J_{22}^\text{IMM}[f_2,f_2]\to K_{22}[f_2,f_2]$. With this change, the nonzero elements of $P_{2,ij}^*$ are \cite{AGG25}
\beq
\label{5.8}
P_{2,yy}^*=P_{2,zz}^*=\frac{1+\al_{22}}{1+\al_{22}+\epsilon_{22}^*}=\frac{2}{3-\al_{22}}, \quad P_{2,xx}^*=d-(d-1)P_{2,yy}^*=\frac{d}{2}\frac{d+3-(d+1)\al_{22}}{d+1-\al_{22}},
\eeq
\beq
\label{5.9}
P_{2,xy}^*=-\frac{2P_{2,yy}^*}{1+\al_{22}+2\epsilon_{22}^*}a^*=-\frac{8a^*}{(3-\al_{22})^2(1+\al_{22})}, \quad
\eeq
\beq
a^*=\frac{a}{\nu_{22}}=-\frac{d}{2}\frac{\zeta_2^*}{P_{2,xy}^*}=
\frac{(3-\al_{22})(1+\al_{22})}{8}\sqrt{d(1-\al_{22})}.
\eeq

\begin{figure}[t]
\begin{center}
\begin{tabular}{lr}
\resizebox{10.0cm}{!}{\includegraphics{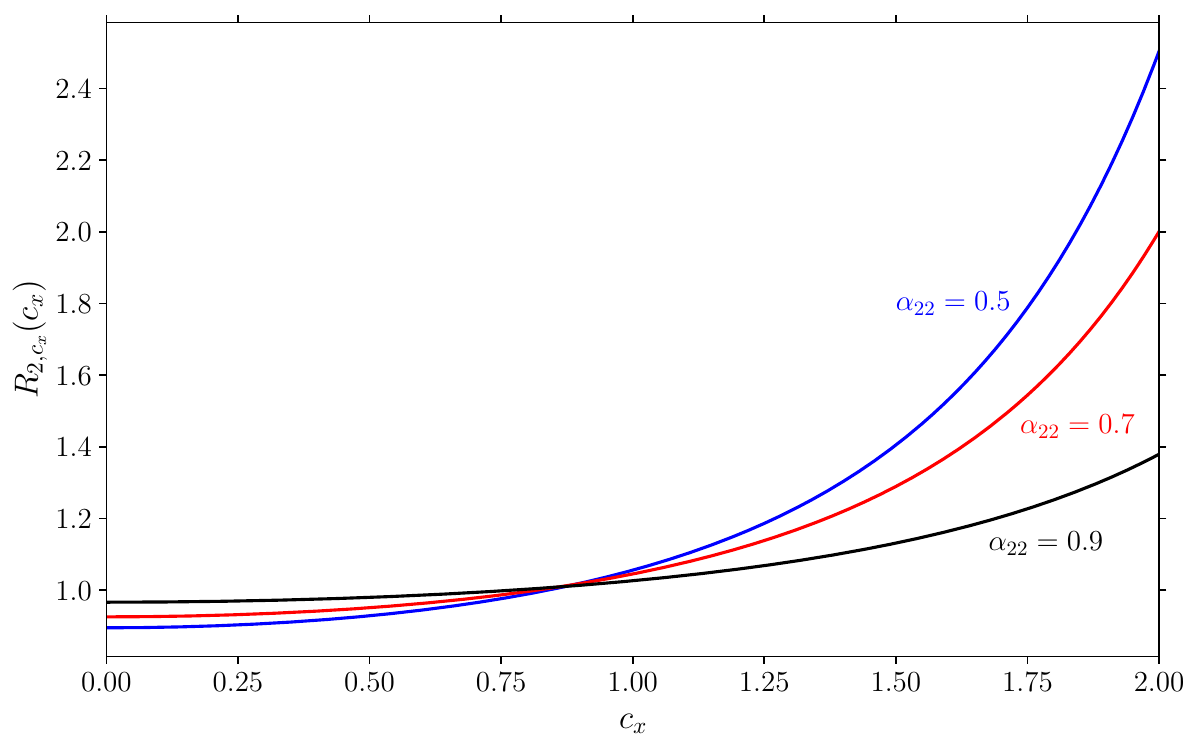}}
\end{tabular}
\end{center}
\caption{Plot of the ratio $R_{2,c_x}(c_x)=\varphi_{2,x}(c_x)/\varphi_{2,x}^{\text{el}}(c_x)$ obtained by means of the VGS kinetic model vs the (scaled) velocity $c_x$ for three different values of the coefficient of restitution $\alpha_{22}$: $\alpha_{22} = 0.9$, $0.7$, and $0.5$.
\label{figfdv2}}
\end{figure}

As said in the Introduction, one of the main advantages of considering a kinetic model instead of the true Boltzmann equation is the possibility of obtaining the velocity distribution function. In the case of the excess granular gas, the kinetic equation obeying its distribution function $f_2(\mathbf{V})$ can be cast into the form
\beq
\label{5.9.1}
\Big(\omega_{22}^*-\epsilon_{22}^*\mathbf{V}\cdot \frac{\partial}{\partial \mathbf{V}}-2a^*V_y \frac{\partial}{\partial V_x}\Big)f_2=\Phi_2(\mathbf{V}),
\eeq
where $\omega_{22}^*=\omega_{22}/\nu_{22}=1+\al_{22}-3\epsilon_{22}^*$, $\epsilon_{22}^*=\epsilon_{22}/\nu_{22}=(1-\al_{22}^2)/4$,
and $\Phi_2\equiv (1+\al_{22})f_{22}$. A formal (hydrodynamic) solution to equation \eqref{5.9.1} is
\begin{align}
\label{5.9.2}
f_2(\mathbf{V})  =\Big(\omega_{22}^*-\epsilon_{22}^*\mathbf{V}\cdot \frac{\partial}{\partial \mathbf{V}}-2a^*V_y \frac{\partial}{\partial V_x}\Big)^{-1}\Phi_2(\mathbf{V}) =\int_0^\infty\; \dd s \; e^{-\omega_{22}^* s} \exp \Big(\epsilon_{22}^* s \mathbf{V}\cdot \frac{\partial}{\partial \mathbf{V}}+2a^*V_y \frac{\partial}{\partial V_x}\Big)\Phi_2(\mathbf{V}).
\end{align}
The action of the shift operators $e^{\epsilon_{22}^* s \mathbf{V}\cdot \frac{\partial}{\partial \mathbf{V}}}$ and $e^{2a^*V_y \frac{\partial}{\partial V_x}}$ in velocity space on an arbitrary function of velocity $g(\mathbf{V})$ is
\beq
\label{5.9.3}
e^{\epsilon_{22}^* s \mathbf{V}\cdot \frac{\partial}{\partial \mathbf{V}}}g(\mathbf{V})=g\Big(e^{\epsilon_{22}^* s}V_x, e^{\epsilon_{22}^* s}V_y, e^{\epsilon_{22}^* s}V_z\Big), \quad e^{2a^*V_y \frac{\partial}{\partial V_x}}g(\mathbf{V})=g\Big(V_x+2a^* sV_y, V_y, V_z\Big),
\eeq
where the three-dimensional case ($d=3$) has been considered for the sake of concreteness. Taking into account the relations \eqref{5.9.3}, the distribution $f_2(\mathbf{V})$ can be written as
$f_2(\mathbf{V})=n_2 (2T_2/m_2)^{-d/2} \varphi_2(\mathbf{c})$, where $\mathbf{c}=\mathbf{V}/\sqrt{2T_2/m_2}$ is a dimensionless velocity and
\beq
\label{5.9.4}
\varphi_2(\mathbf{c})=\pi^{-d/2}(1+\al_{22})\int_0^\infty\; \dd s \; e^{-\omega_{22}^* s} \exp \Big[
- e^{2\epsilon_{22}^* s}\left(c^2+4a^* s c_x c_y+4 a^{*2}s^2 c_y^2\right)\Big].
\eeq

\noindent The dependence of the distribution $\varphi_2(\mathbf{c})$ on the coefficient of restitution $\alpha_{22}$ provided by equation \eqref{5.9.4} has been shown to agree very well (at least in the regime of thermal velocities, i.e., $|\mathbf{c}|\sim 1$) with DSMC results \cite{BRM97}.

To illustrate the dependence of $\varphi_2(\mathbf{c})$ on the dimensionless velocity $\mathbf{c}$, we consider $d=3$ and define the marginal distribution
\beq
\label{5.9.5}
\varphi_{2,x}(c_x)=\int_{-\infty}^{+\infty} \dd c_y \int_{-\infty}^{+\infty} \dd c_z \; \varphi_2(\mathbf{c}).
\eeq
Substituting equation \eqref{5.9.4} into equation \eqref{5.9.5} and performing the velocity integrals one gets
\beq
\label{5.9.6}
\varphi_{2,x}(c_x)=\pi^{-1/2}(1+\al_{22})\int_0^\infty\; \dd s \; \frac{e^{-(1+\al_{22}-\epsilon_{22}^*) s}}
{\sqrt{1+4a^{*2}s^2}}
\exp \Big(- e^{2\epsilon_{22}^* s}\frac{c_x^2}{1+4a^{*2}s^2}\Big).
\eeq
For elastic collisions ($\al_{22}=1$), $\epsilon_{22}^*=0$, $a^*=0$, and $\varphi_{2,x}(c_x)$ reduces to the equilibrium distribution $\varphi_{2,x}^{\text{el}}(c_x)=\pi^{-1/2} e^{-c_x^2} $. Figure \ref{figfdv2} plots the ratio $R_{2,c_x}(c_x)=\varphi_{2,x}(c_x)/\varphi_{2,x}^{\text{el}}(c_x)$ as a function of the reduced velocity $c_x$ for three different values of $\alpha_{22}$.
We observe a quite distortion of the scaled USF distribution $\varphi_{2,x}(c_x)$ from its equilibrium value $\varphi_{2,x}^{\text{el}}$ since the ratio $R_{2,c_x}(c_x)$ deviates clearly from 1. The deviation of $\varphi_{2,x}(c_x)$ from $\varphi_{2,x}^{\text{el}}$ increases with increasing dissipation as expected.

\begin{figure}[t]
\begin{center}
\begin{tabular}{lr}
\resizebox{10.0cm}{!}{\includegraphics{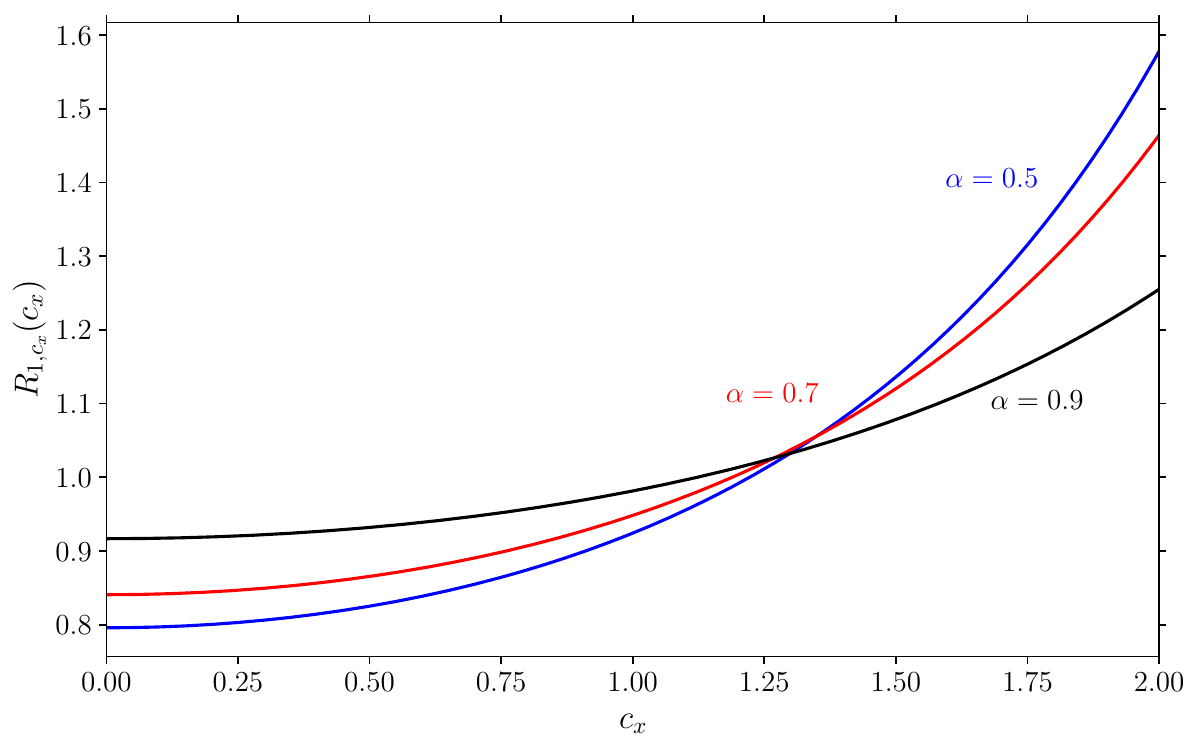}}
\end{tabular}
\end{center}
\caption{Plot of the ratio $R_{1,c_x}(c_x)=\varphi_{1,x}(c_x)/\varphi_{1,x}^{\text{el}}(c_x)$ obtained by means of the VGS kinetic model vs the (scaled) velocity $c_x$ for $m_1/m_2 = \sigma_1/\sigma_2 = 0.5$, and three different values of the (common)
coefficient of restitution $\alpha_{22} =\alpha_{12} = \alpha$: $\alpha = 0.9$, $0.7$, and $0.5$.
\label{figfdv1}}
\end{figure}

\subsection{Tracer particles}

To determine the rheological properties of the tracer particles from the VGS model one has to make the replacement $J_{12}^\text{IMM}[f_1,f_2]\to K_{12}[f_1,f_2]$ in the second equation of \eqref{3.3}. With this change, the expressions of the nonzero elements of $P_{1,ij}^*$ are \cite{AGG25}
\beq
\label{5.10}
P_{1,yy}^*=\frac{2}{(1+\al_{12})\nu_{12}^*+2\epsilon_{12}^*}, \quad P_{1,xy}^*=-\frac{2a^*}{(1+\al_{12})\nu_{12}^*+2\epsilon_{12}^*}P_{1,yy}^*,
\quad P_{1,xx}^*=\gamma-(d-1)P_{1,yy}^*.
\eeq
In equations\ \eqref{5.9} and \eqref{5.10} we have introduced the quantities
\beq
\label{5.10.1}
\nu_{12}^*=\frac{\nu_{12}}{\nu_{22}}=\left(\frac{\sigma_{12}}{\sigma_2}\right)^{d-1}\sqrt{\frac{1+\theta}{2\theta}}, \quad
\epsilon_{12}^*=\frac{\epsilon_{12}}{\nu_{22}}=\frac{1}{2}\nu_{12}^*\mu_{21}^2(1+\theta)(1-\al_{12}^2).
\eeq
The temperature ratio $\gamma=T_1/T_2$ can be obtained from the relationship \eqref{3.11} where $\zeta_2^*$ must be replaced by $\epsilon_{22}^*$ and $\zeta_1^*$ is also given by Eq.\ \eqref{3.12}.

As in the case of the excess granular gas, one can obtain the explicit form of the velocity distribution function $f_1(\mathbf{V})$. By following similar mathematical steps as those made before for the distribution $f_2(\mathbf{V})$, the distribution $f_1(\mathbf{V})$ for $d=3$ can be written as $f_1(\mathbf{V})=n_1 (2T_2/m_2)^{-d/2} \varphi_1(\mathbf{c})$, where the scaled distribution $\varphi_1(\mathbf{c})$ is
\beq
\label{5.10.2}
\varphi_1(\mathbf{c})=\pi^{-d/2}(1+\al_{12})\nu_{12}^*\theta_{12}^{d/2}\int_0^\infty\; \dd s \; e^{-\omega_{12}^* s} \exp \Big[-\theta_{12}
e^{2\epsilon_{12}^* s}\left(c^2+4a^* s c_x c_y+4 a^{*2}s^2 c_y^2\right)\Big],
\eeq
where $\omega_{12}^*=\omega_{12}/\nu_{22}=(1+\al_{22})\nu_{12}^*-3\epsilon_{12}^*$, $\theta_{12}=\mu/\gamma_{12}$, and $\gamma_{12}=T_{12}/T_2=\gamma+2\mu_{12}\mu_{21}(1-\gamma)$. Here, $\mu=m_1/m_2$ is the mass ratio. The corresponding marginal distribution
\beq
\label{5.10.3}
\varphi_{1,x}(c_x)=\int_{-\infty}^{+\infty} \dd c_y \int_{-\infty}^{+\infty} \dd c_z \; \varphi_1(\mathbf{c})
\eeq
for $d=3$ is
\beq
\label{5.10.4}
\varphi_{1,x}(c_x)=\pi^{-1/2}(1+\al_{12})\nu_{12}^*\theta_{12}^{1/2}\int_0^\infty\; \dd s \; \frac{e^{-[(1+\al_{12})\nu_{12}^*-\epsilon_{12}^*] s}}
{\sqrt{1+4a^{*2}s^2}}
\exp \Big(- \theta_{12}e^{2\epsilon_{12}^* s}\frac{c_x^2}{1+4a^{*2}s^2}\Big).
\eeq
For elastic collisions ($\al_{22}=\al_{12}=1$), $\epsilon_{12}^*=0$, $a^*=0$, $\gamma=1$, and so $\varphi_{1,x}(c_x)$ reduces to the equilibrium distribution $\varphi_{1,x}^{\text{el}}(c_x)=\pi^{-1/2} \mu^{1/2} e^{-\mu c_x^2} $. The dependence of the ratio $R_{1,c_x}(c_x)=\varphi_{1,x}(c_x)/\varphi_{1,x}^{\text{el}}(c_x)$ on $c_x$ is plotted in figure \ref{figfdv1} for $m_1/m_2 = \sigma_1/\sigma_2 = 0.5$. As for the granular gas, we observe that the deviation of $R_{1,c_x}(c_x)$ from 1 is more noticeable as the inelasticity in collisions increases.

In the sheared diffusion problem, the first-order distribution function $f_1^{(1)}$ obeys the kinetic equation \eqref{4.7.1}, except that $J_{12}^\text{IMM}[f_1^{(0)},f_2^{(1)}]\to K_{12}[f_1^{(0)},f_2^{(1)}]=0$ and the Boltzmann--Lorentz collision operator $J_{12}^\text{IMM}[f_1^{(1)},f_2^{(0)}]$ must be replaced by the operator
\beq
\label{5.11}
K_{12}[f_1^{(1)},f_2^{(0)}]=-\frac{1+\al_{12}}{2}\nu_{12}\Big(f_1^{(1)}-f_{12}^{(1)}\Big)+\frac{\epsilon_{12}}{2}\Big(
\frac{\partial}{\partial \mathbf{c}}\cdot \mathbf{c}f_1^{(1)}-\frac{\mathbf{j}_1^{(1)}}{\rho_1}\cdot \frac{\partial}{\partial \mathbf{c}} f_1^{(0)}\Big),
\eeq
where
\beq
\label{5.12}
f_{12}^{(1)}(\mathbf{c})=\frac{\mu_{12}}{n_1T_{12}}\mathbf{c}\cdot \mathbf{j}_1^{(1)}f_{12}^{(0)}(\mathbf{c}), \quad
f_{12}^{(0)}(\mathbf{c})=n_1\left(\frac{m_1}{2\pi
T_{12}}\right)^{d/2} \exp\left(-\frac{m_1}{2T_{12}}c^2\right).
\eeq
With these replacements, it is easy to see that the diffusion transport coefficients of the VGS kineitc model are the solutions of the algebraic equations \eqref{4.19}--\eqref{4.21} except that $\Omega$ must be replaced by
\beq
\label{5.14}
\Omega'=\frac{1+\al_{12}}{2}\mu_{21}\nu_{12}.
\eeq
Note that in equations \eqref{4.19}--\eqref{4.21} the derivatives of the pressure tensors $P_{2,k \ell}^*$ and $P_{1,k \ell}^*$ with respect to $a^*$ in the steady state obtained from the kinetic model differ from those obtained from IMM and IHS. These derivatives are also displayed in the Appendix \ref{appB}.


\section{Comparison with the Boltzmann results for IHS}
\label{sec6}

\begin{figure}[t]
\begin{center}
\begin{tabular}{lr}
\resizebox{10.0cm}{!}{\includegraphics{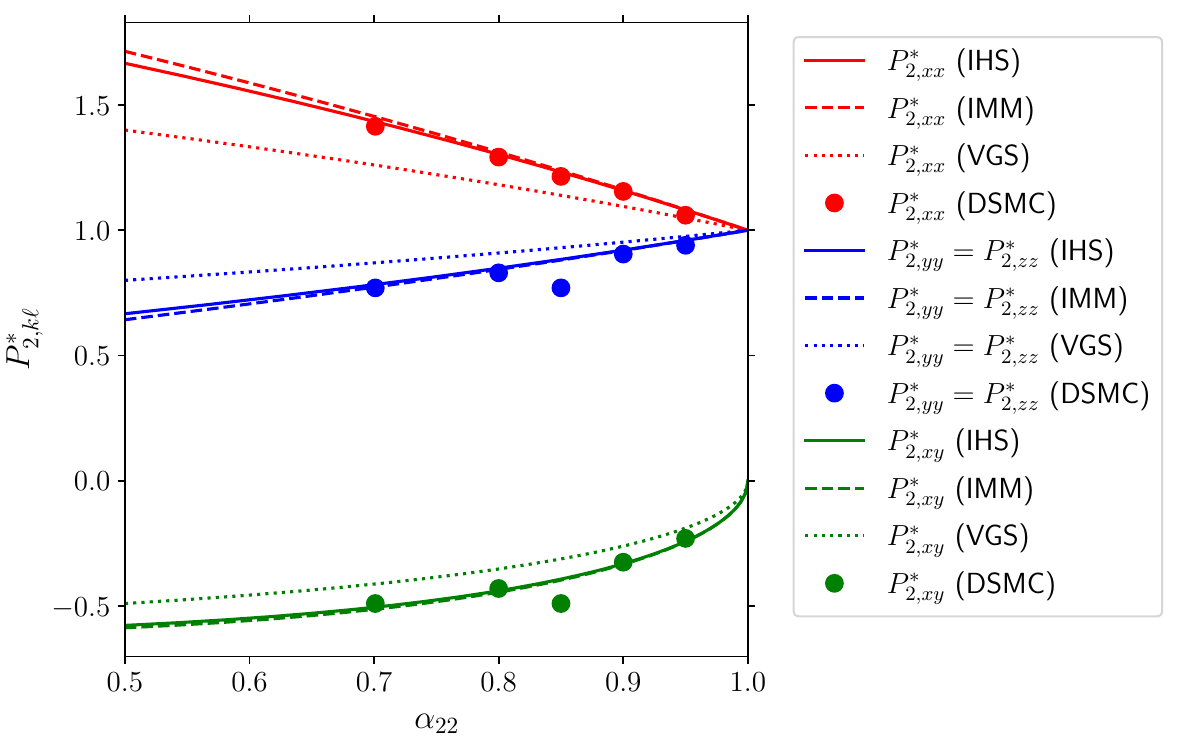}}
\end{tabular}
\end{center}
\caption{Plot of the (reduced) elements of the pressure tensor $P_{2,k\ell}^*$ as functions of the coefficient of restitution $\al_{22}$ for a three-dimensional single gas. The solid lines are the approximate results derived for IHS from the leading Sonine approximation, the dashed lines correspond to the results obtained for IMM, and the dotted lines refer to the results of the VGS kinetic model. Symbols are the Monte Carlo simulations for IHS obtained in the paper \cite{MG02a}.
\label{fig1}}
\end{figure}

\begin{figure}[h]
\begin{center}
\begin{tabular}{lr}
\resizebox{10.0cm}{!}{\includegraphics{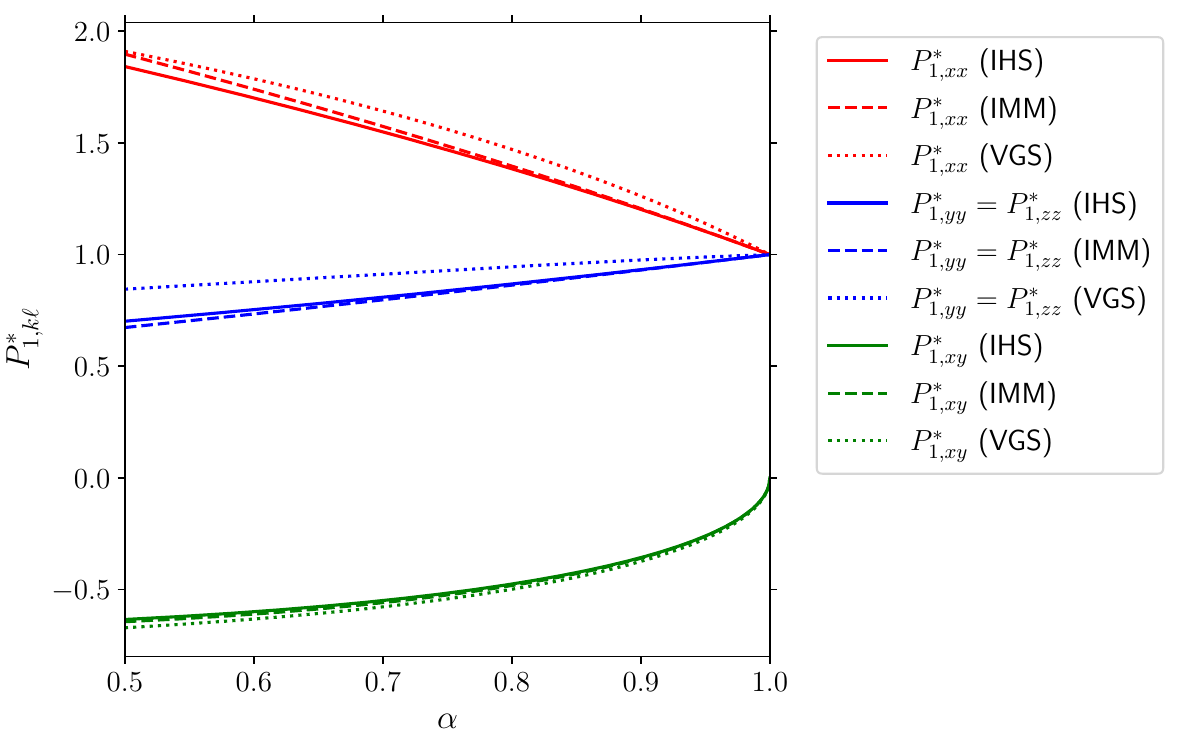}}
\end{tabular}
\end{center}
\caption{Plot of the (reduced) elements of the pressure tensor $P_{1,k\ell}^*$ as functions of the (common) coefficient of restitution $\al_{22}=\al_{12}\equiv \al$ for a three-dimensional single system ($d=3$) in the case $\sigma_1/\sigma_2=m_1/m_2=0.5$. The solid lines are the approximate results derived for IHS from the leading Sonine approximation, the dashed lines correspond to the results obtained for IMM, and the dotted lines refer to the results of the VGS kinetic model.
\label{fig2}}
\end{figure}

\begin{figure}[h]
\begin{center}
\begin{tabular}{lr}
\resizebox{10.0cm}{!}{\includegraphics{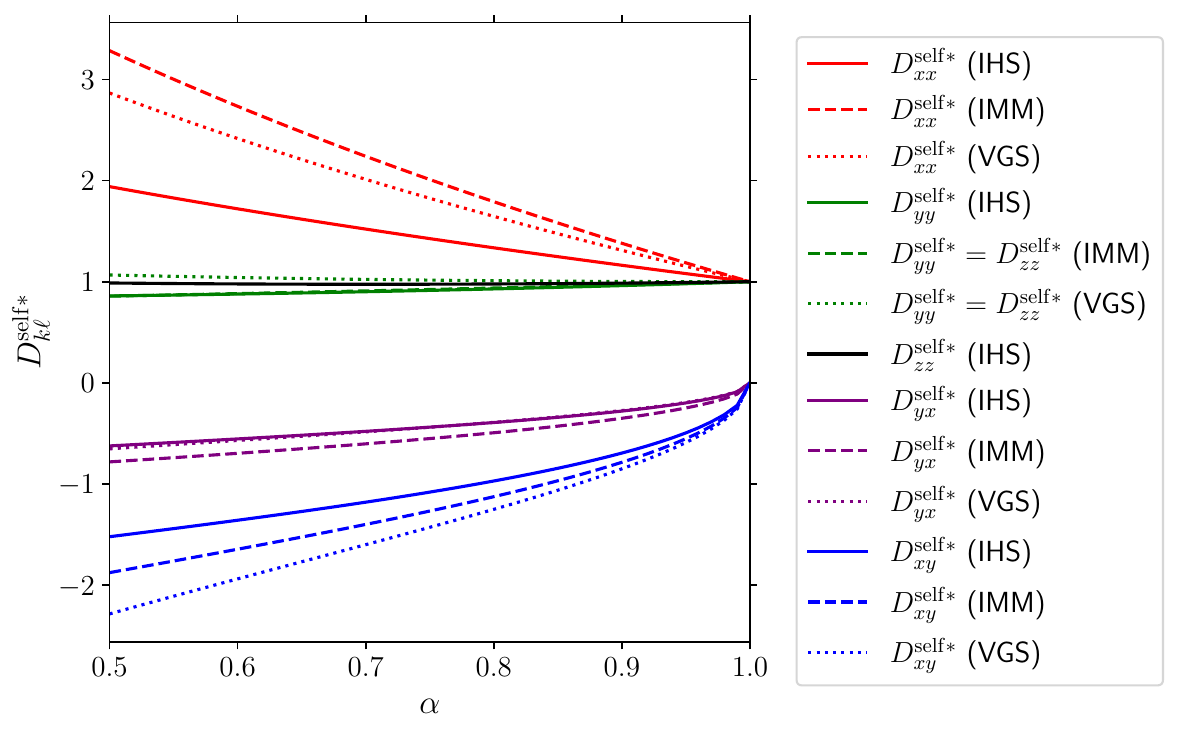}}
\end{tabular}
\end{center}
\caption{Plot of the (reduced) elements of the self-diffusion tensor $D_{xx}^{\text{self}*}$, $D_{yy}^{\text{self}*}$, $D_{zz}^{\text{self}*}$, $D_{xy}^{\text{self}*}$, and $D_{yx}^{\text{self}*}$ as functions of the (common) coefficient of restitution $\al_{22}=\al_{12}=\al$ for a three-dimensional system in the case $\sigma_1/\sigma_2=m_1/m_2=1$. The solid lines are the approximate results derived for IHS from the leading Sonine approximation, the dashed lines correspond to the results obtained for IMM, and the dotted lines refer to the results of the VGS kinetic model. Note that $D_{p,yy}^{\text{self}*}=D_{p,zz}^{\text{self}*}$ in the results obtained from IMM and the kinetic model. \label{self}}
\end{figure}

\begin{figure}[h!]
\begin{center}
\begin{tabular}{lr}
\resizebox{10.0cm}{!}{\includegraphics{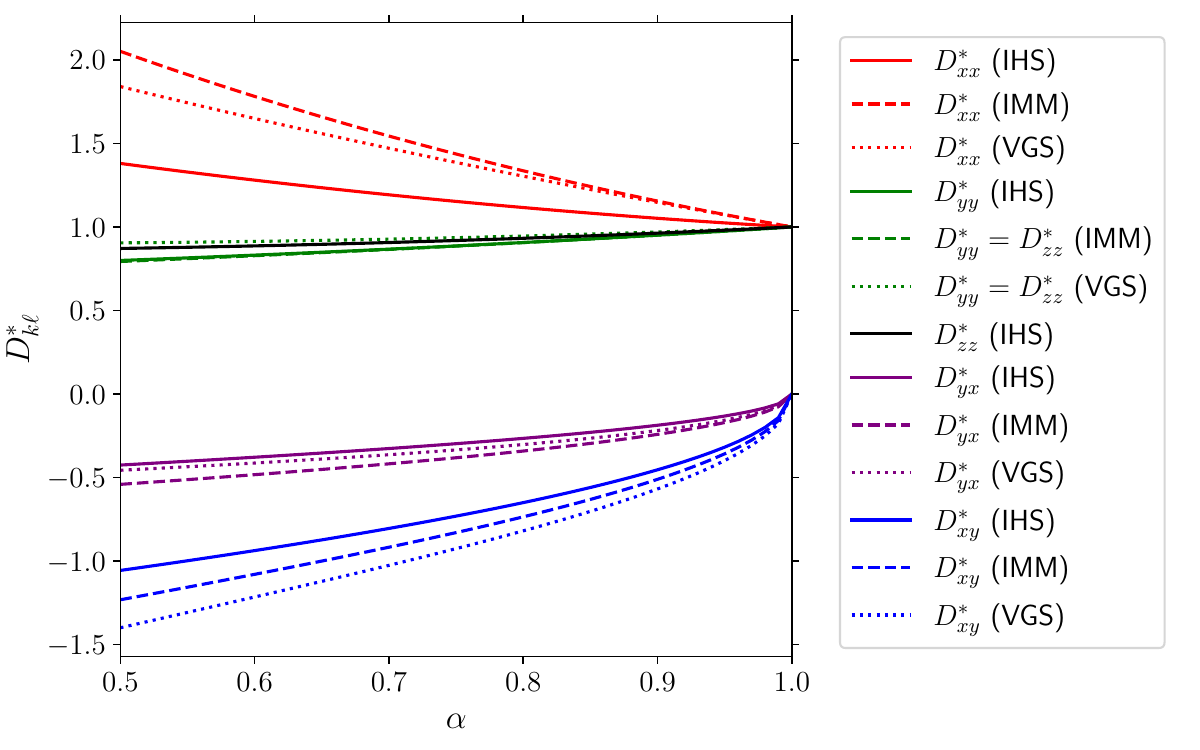}}
\end{tabular}
\end{center}
\caption{Plot of the (reduced) elements $D_{xx}^*$, $D_{yy}^*$, $D_{zz}^*$, $D_{xy}^*$, and $D_{yx}^*$ as functions of the (common) coefficient of restitution $\al$ for a three-dimensional system in the case $\sigma_1/\sigma_2=1$ and $m_1/m_2=0.5$. The solid lines are the approximate results derived for IHS from the leading Sonine approximation, the dashed lines correspond to the results obtained for IMM, and the dotted lines refer to the results of the VGS kinetic model. Note that $D_{yy}^*=D_{zz}^*$ in the results obtained from IMM and the kinetic model.
\label{fig3}}
\end{figure}
\begin{figure}[t]
\begin{center}
\begin{tabular}{lr}
\resizebox{10.0cm}{!}{\includegraphics{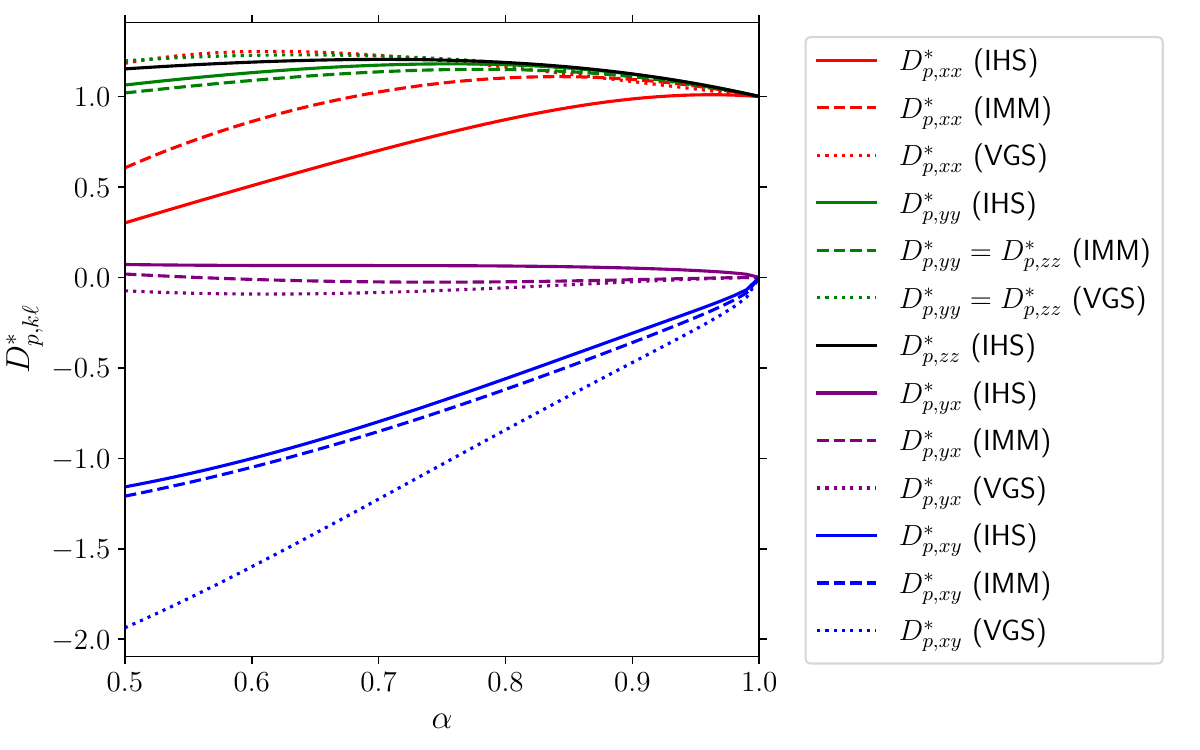}}
\end{tabular}
\end{center}
\caption{Plot of the (reduced) elements $D_{p,xx}^*$, $D_{p,yy}^*$, $D_{p,zz}^*$, $D_{p,xy}^*$, and $D_{p,yx}^*$ as functions of the (common) coefficient of restitution $\al$ for a three-dimensional system in the case $\sigma_1/\sigma_2=1$ and $m_1/m_2=0.5$. The solid lines are the approximate results derived for IHS from the leading Sonine approximation, the dashed lines correspond to the results obtained for IMM, and the dotted lines refer to the results of the VGS kinetic model. Note that $D_{p,yy}^*=D_{p,zz}^*$ in the results obtained from IMM and the kinetic model. \label{fig4}}
\end{figure}
\begin{figure}[h!]
\begin{center}
\begin{tabular}{lr}
\resizebox{10.0cm}{!}{\includegraphics{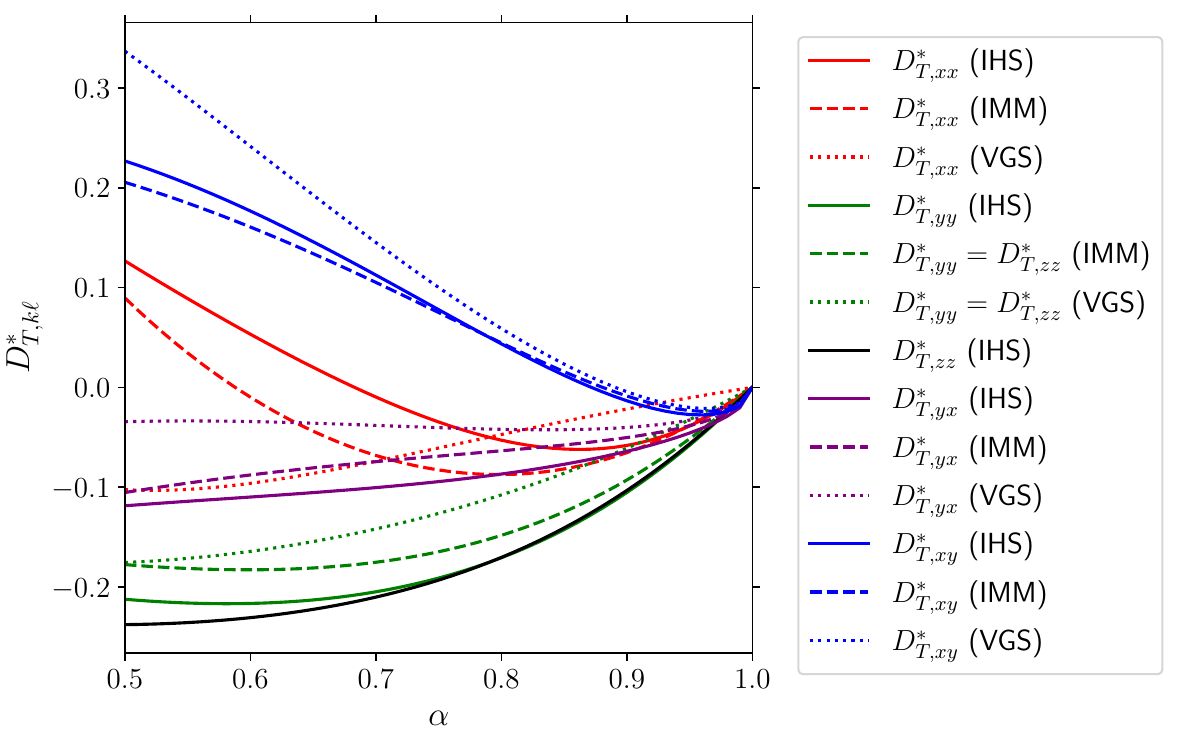}}
\end{tabular}
\end{center}
\caption{Plot of the (reduced) elements $D_{T,xx}^*$, $D_{T,yy}^*$, $D_{T,zz}^*$, $D_{T,xy}^*$, and $D_{T,yx}^*$ as functions of the (common) coefficient of restitution $\al$ for a three-dimensional system in the case $\sigma_1/\sigma_2=1$ and $m_1/m_2=0.5$. The solid lines are the approximate results derived for IHS from the leading Sonine approximation, the dashed lines correspond to the results obtained for IMM, and the dotted lines refer to the results of the VGS kinetic model. Note that $D_{T,yy}^*=D_{T,zz}^*$ in the results obtained from IMM and the kinetic model. \label{fig5}}
\end{figure}

In this section, we want to compare the results obtained from both the Boltzmann equation for IMM and the VGS kinetic model with those previously derived from the IHS \cite{G02,G07}. In dimensionless form, the rheological properties and the diffusion coefficients of the system (granular gas plus tracer particles) depend on five quantities: the diameter $\sigma_1/\sigma_2$ and mass $m_1/m_2$ ratios, the coefficients of restitution $\al_{22}$ and $\al_{12}$, and the dimensionality $d$ of the system. Thus, given that the parameter space of the system  is large, henceforth we consider a three-dimensional gas ($d=3$) and a common coefficient of restitution $\al_{22}=\al_{12}\equiv \al$. This reduces the parameter space to three quantities: $\sigma_1/\sigma_2$, $m_1/m_2$, and $\al$.

\subsection{Rheological properties}

First, we consider the rheological properties. In the case of the excess granular gas, figure \ref{fig1} shows $P_{2,k\ell}^*$ versus $\al_{22}$. We have also included computer simulation results \cite{MG02a} obtained by numerically solving the (inelastic) Boltzmann equation for IHS by means of the direct simulation Monte Carlo (DSMC) method \cite{B94}. Comparing the three approaches, we see that the quantitative discrepancies between the VGS model's theoretical predictions and the approximate IHS results are larger than those found for the IMM results. In fact, the theoretical results for IHS and IMM disagree very little; this difference increases slightly with inelasticity. Additionally, we observe excellent agreement between the Boltzmann theory for both interaction models and Monte Carlo simulations, even in the case of strong dissipation. Similar conclusions are reached for the reduced pressure tensor, $P_{1,k\ell}^*$. Figure \ref{fig2} illustrates this behavior, showing the dependence of $P_{1,k\ell}^*$ on $\al$ for the case $\sigma_1/\sigma_2=1$ and $m_1/m_2=0.5$. The good agreement between the IHS and IMM results is apparent again, especially in the case of the shear stress $P_{1,xy}^*$, which is the most relevant rheological property in a shear flow problem as it defines the non-Newtonian shear viscosity. Although the VGS model predictions are good qualitatively, they exhibit larger discrepancies with the IHS results than the IMM predictions.

\subsection{Tracer diffusion tensors}

We analyze now the dependence of the diffusion coefficients $T_{k\ell}\equiv \left\{D_{k\ell}, D_{p,k\ell},D_{T,k\ell}\right\}$ on the (common) coefficient of restitution $\al$. According to the results derived in sections \ref{sec4} and \ref{sec5}, $T_{xz}=T_{zx}=T_{yz}=T_{zy}=0$ in agreement with the symmetry of the shearing field applied to the system. Thus, in a three-dimensional gas, there are five relevant elements of the tensors $T_{k\ell}$: the three diagonal ($T_{xx}$, $T_{yy}$, and $T_{zz}$) and two off-diagonal elements ($T_{xy}$ and $T_{yx}$). The results obtained here for IMM and from the kinetic model show that in general $T_{xx}\neq T_{yy}=T_{zz}$ and $T_{xy}\neq T_{yx}$. However, in the case of IHS, the approximate results derived in the paper \cite{G07} show that $T_{yy}\neq T_{zz}$ although the difference between both elements is in general very small. As previously mentioned, the diagonal elements, $T_{kk}$, can be considered generalizations of the conventional Navier-Stokes diffusion transport coefficients because they reduce to the conventional coefficients for elastic gases (which is equivalent to zero shear rate in the steady USF state). The off-diagonal elements, $T_{xy}$ and $T_{yx}$, can be seen as generalizations of the Burnett transport coefficients because they account for cross transport effects appearing in far-from-equilibrium states.

Given that we are interested in this paper to assess the dependence of the coefficients $T_{k\ell}$ on inelasticity, we have scaled them with respect to their elastic values except in the case of $D_{T,k\ell}$ since this coefficient vanishes for elastic collisions. Thus, we define the dimensionless coefficients $D_{k\ell}^*=D_{k\ell}/D^\text{el}$ and $D_{p,k\ell}^*=D_{p,k\ell}/D_p^\text{el}$ while $D_{T,k\ell}^*=D_{T,k\ell}/(x_1 T/m_2\nu)$. The effective collision frequency $\nu=\nu_{\text{M},22}$ for IMM, $\nu=\nu_{22}$ for the VGS model, and $\nu=(2/(d+2))\nu_{\text{M},22}$ for IHS.

\begin{figure}[h!]
\begin{center}
\begin{tabular}{lr}
\resizebox{10.0cm}{!}{\includegraphics{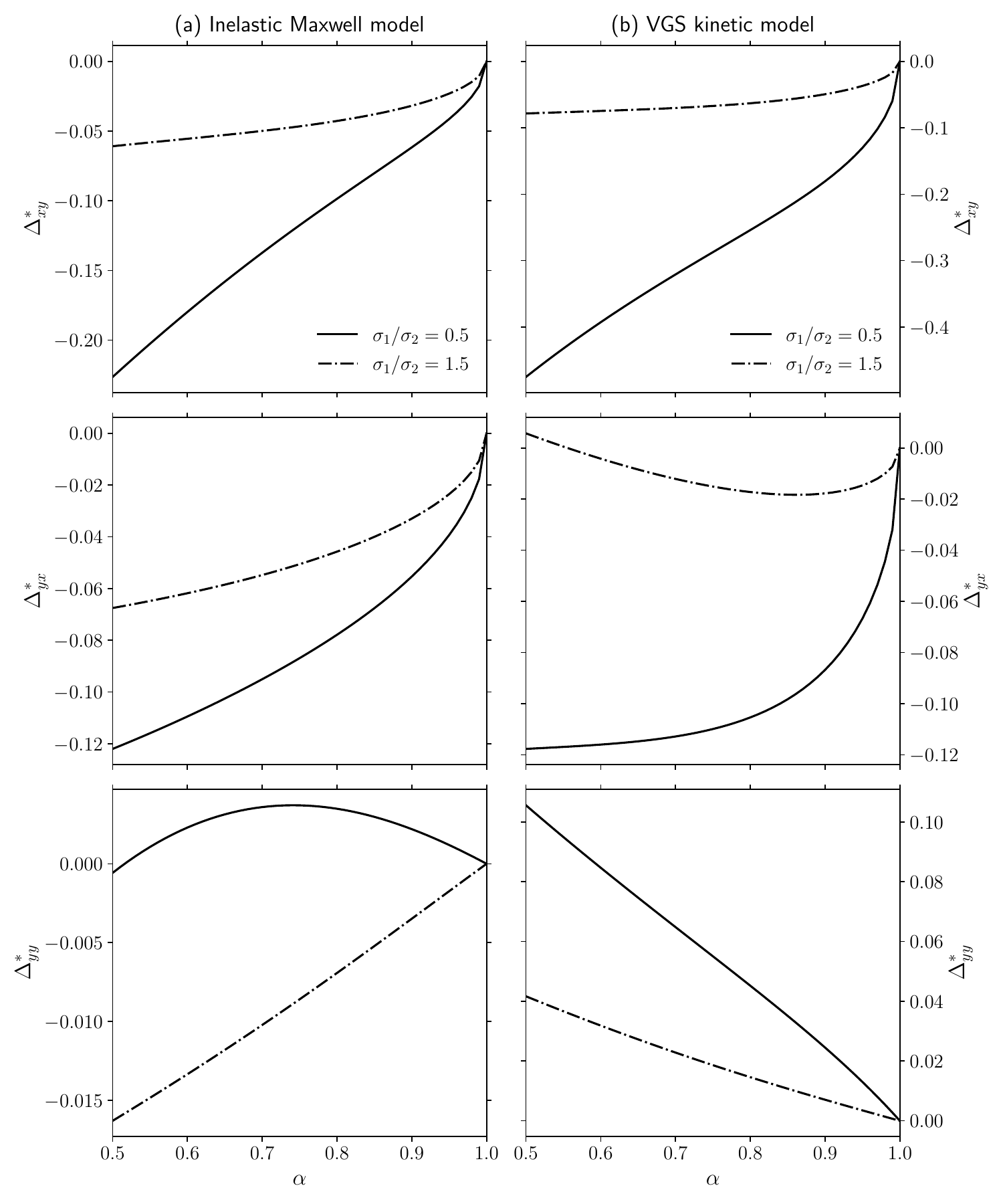}}
\end{tabular}
\end{center}
\caption{Panel (a): Plot of the dimensionless quantities $\Delta_{ij}^*=D_{ij}^{\text{*IMM}}-D_{ij}^{*\text{IHS}}$ as functions of the (common) coefficient of restitution $\al$ for $m_1/m_2=0.25$ and two different values of the diameter ratio: $\sigma_1/\sigma_2=0.5$ (solid lines) and $\sigma_1/\sigma_2=1.5$ (dash-dotted lines). We have considered the elements $xy$, $yx$ and $yy$ of IMM. Panel (b): The same as for the panel (a) for the results obtained from the VGS kinetic model.
\label{Deltaij}}
\end{figure}
\begin{figure}[h!]
\begin{center}
\begin{tabular}{lr}
\resizebox{10.0cm}{!}{\includegraphics{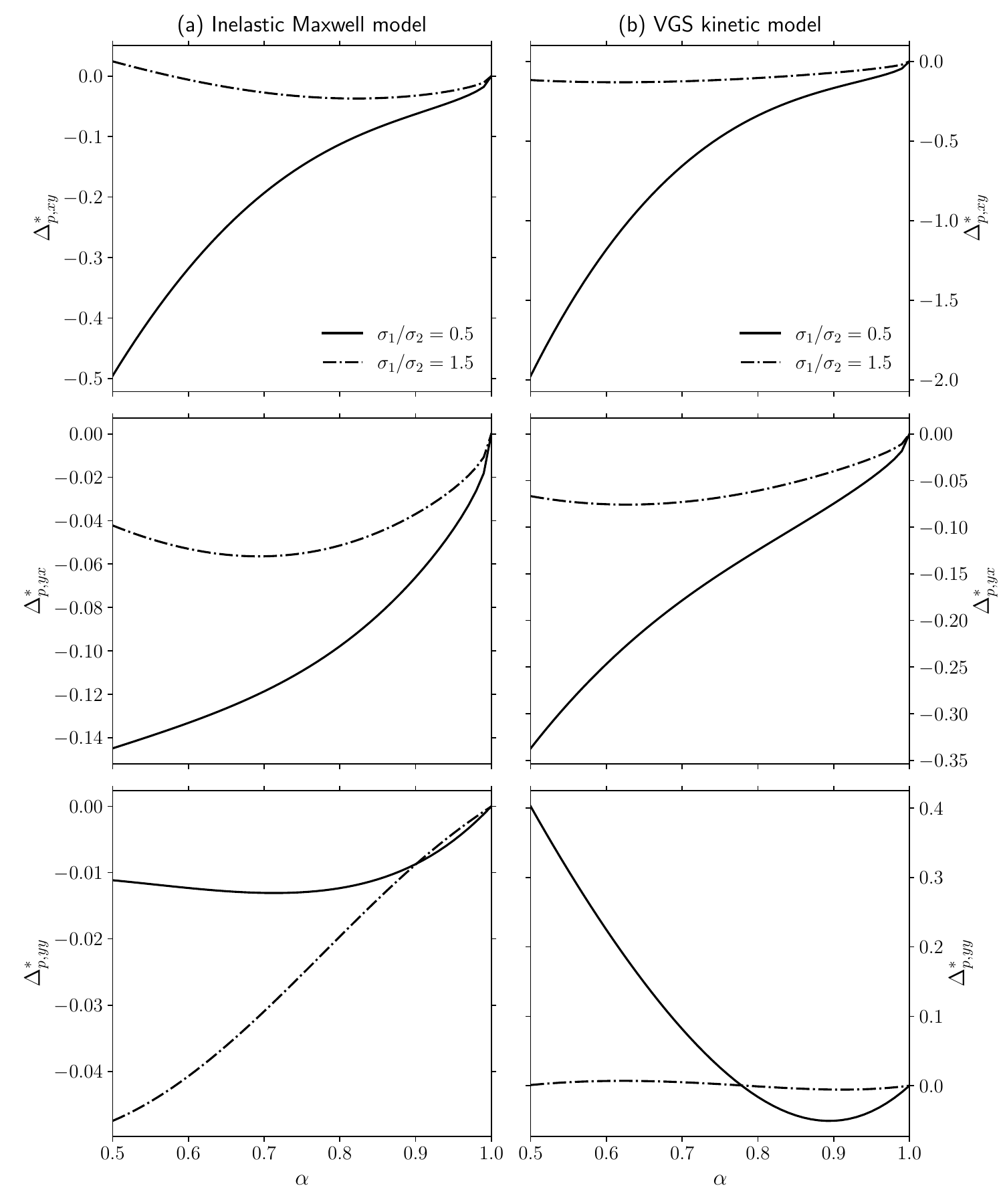}}
\end{tabular}
\end{center}
\caption{Panel (a): Plot of the dimensionless quantities $\Delta_{p,ij}^*=D_{p,ij}^{*\text{IMM}}-D_{p,ij}^{*\text{IHS}}$ as functions of the (common) coefficient of restitution $\al$ for $m_1/m_2=0.25$ and two different values of the diameter ratio: $\sigma_1/\sigma_2=0.5$ (solid lines) and $\sigma_1/\sigma_2=1.5$ (dash-dotted lines). We have considered the elements $xy$, $yx$ and $yy$ of IMM. Panel (b): The same as for the panel (a) for the results obtained from the VGS kinetic model.
\label{Deltapij}}
\end{figure}
\begin{figure}[h!]
\begin{center}
\begin{tabular}{lr}
\resizebox{10.0cm}{!}{\includegraphics{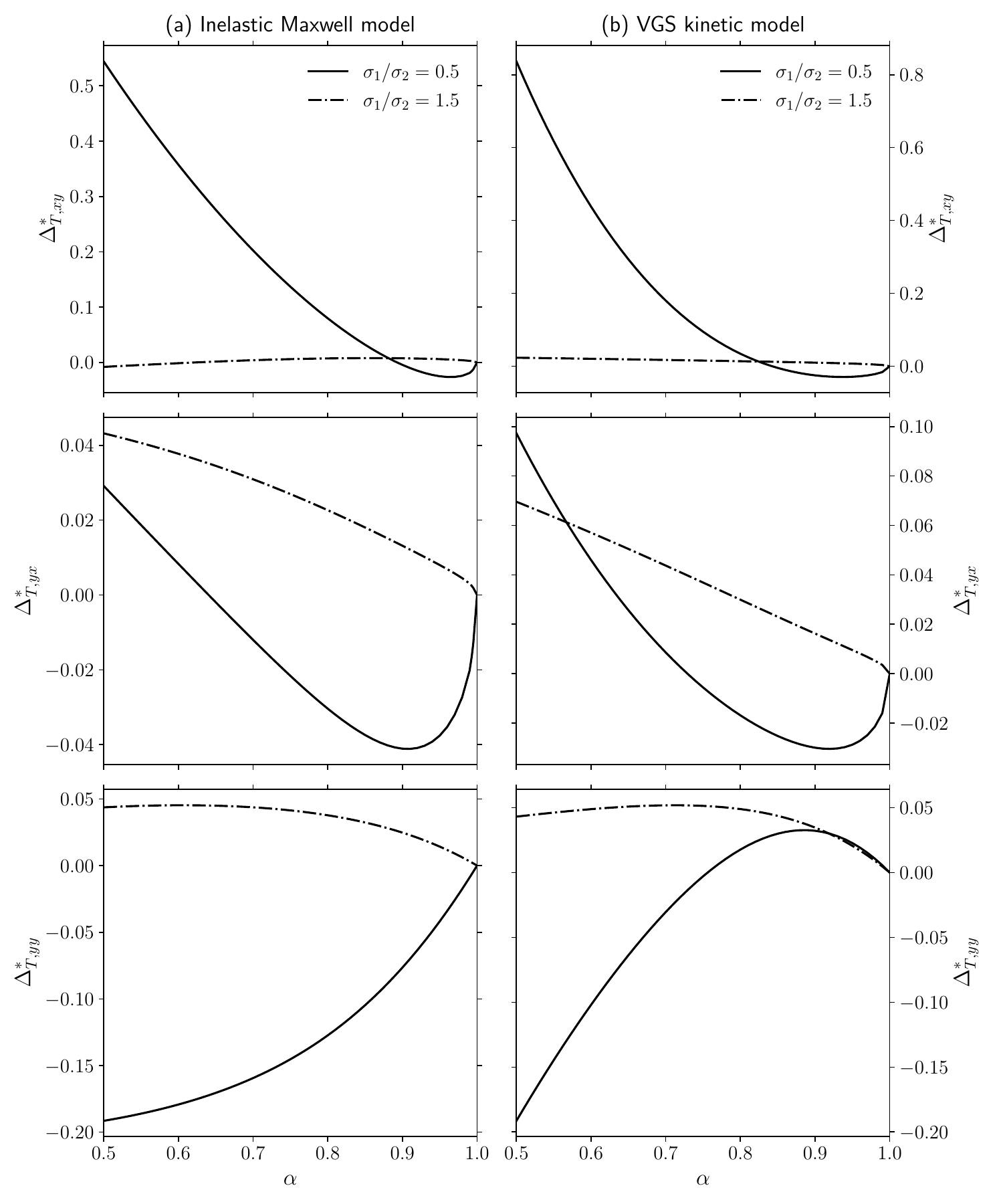}}
\end{tabular}
\end{center}
\caption{Panel (a): Plot of the dimensionless quantities $\Delta_{T,ij}^*=D_{T,ij}^{*\text{IMM}}-D_{T,ij}^{*\text{IHS}}$ as functions of the (common) coefficient of restitution $\al$ for $m_1/m_2=0.25$ and two different values of the diameter ratio: $\sigma_1/\sigma_2=0.5$ (solid lines) and $\sigma_1/\sigma_2=1.5$ (dash-dotted lines). We have considered the elements $xy$, $yx$ and $yy$ of IMM. Panel (b): The same as for the panel (a) for the results obtained from the VGS kinetic model. \label{DeltaTij}}
\end{figure}

Before considering the general case, we will first study the self-diffusion problem, i.e., when the tracer and the particles of the granular gas are mechanically equivalent. As stated in section \ref{sec4}, in this case,  $D_{p,k\ell}=D_{T,k\ell}=0$ and the self-diffusion tensor is given by equation \eqref{4.26.2}. The dependence of the nonzero elements of the (reduced) self-diffusion tensor $D_{k\ell}^{\text{self}*}$ on the (common) coefficient of restitution $\al$ is plotted in figure \ref{self}. The self-diffusion problem involves only single-particle motion and it is therefore somewhat simpler to compute the diffusion coefficients. In particular, the granular temperature $T_2$ is the same as that of the tracer particles $T_1$. We observe that the deviation from the functional form for elastic collisions is in general quite significant, even for moderate dissipation. With respect to the comparison between IHS, IMM and VGS results, we observe good qualitative agreement between the three models. At a quantitative level, however, we find good agreement for the $yy$ and $yx$ elements, but more significant differences are observed for the $xx$ and $xy$ elements. In this latter case, the VGS results are closer to the IHS results than the IMM results for the $xx$ element, but the opposite happens for the $xy$ element.

As previously mentioned, some studies \cite{C97,ALJR21} have performed molecular dynamics simulations to measure the non-zero elements of the self-diffusion tensor. As discussed in  \cite{G02}, the densities analyzed in these simulations generally prevent us from making quantitative comparisons between our theory, which is limited to the low-density regime, and these computer simulations. However, we find that the qualitative dependence of the self-diffusion tensor on dissipation generally agrees with the results derived from IMM and VGS models. These simulations study the $\al$-dependence of the reduced tensor $\widetilde{D}_{k\ell}=D_{k\ell}^{\text{self}*}/a^*$. Kinetic theory and simulations predict that $\widetilde{D}_{xx}>\widetilde{D}_{zz}>\widetilde{D}_{yy}$  while in general the elements $\widetilde{D}_{k\ell}$ decrease with increasing dissipation. However, the simulations show that the value of the off-diagonal element $-\widetilde{D}_{xy}$ is roughly the same magnitude as that of the diagonal element $\widetilde{D}_{yy}$. Our results disagree with this prediction.

\begin{figure}[h!]
\begin{center}
\begin{tabular}{lr}
\resizebox{10.0cm}{!}{\includegraphics{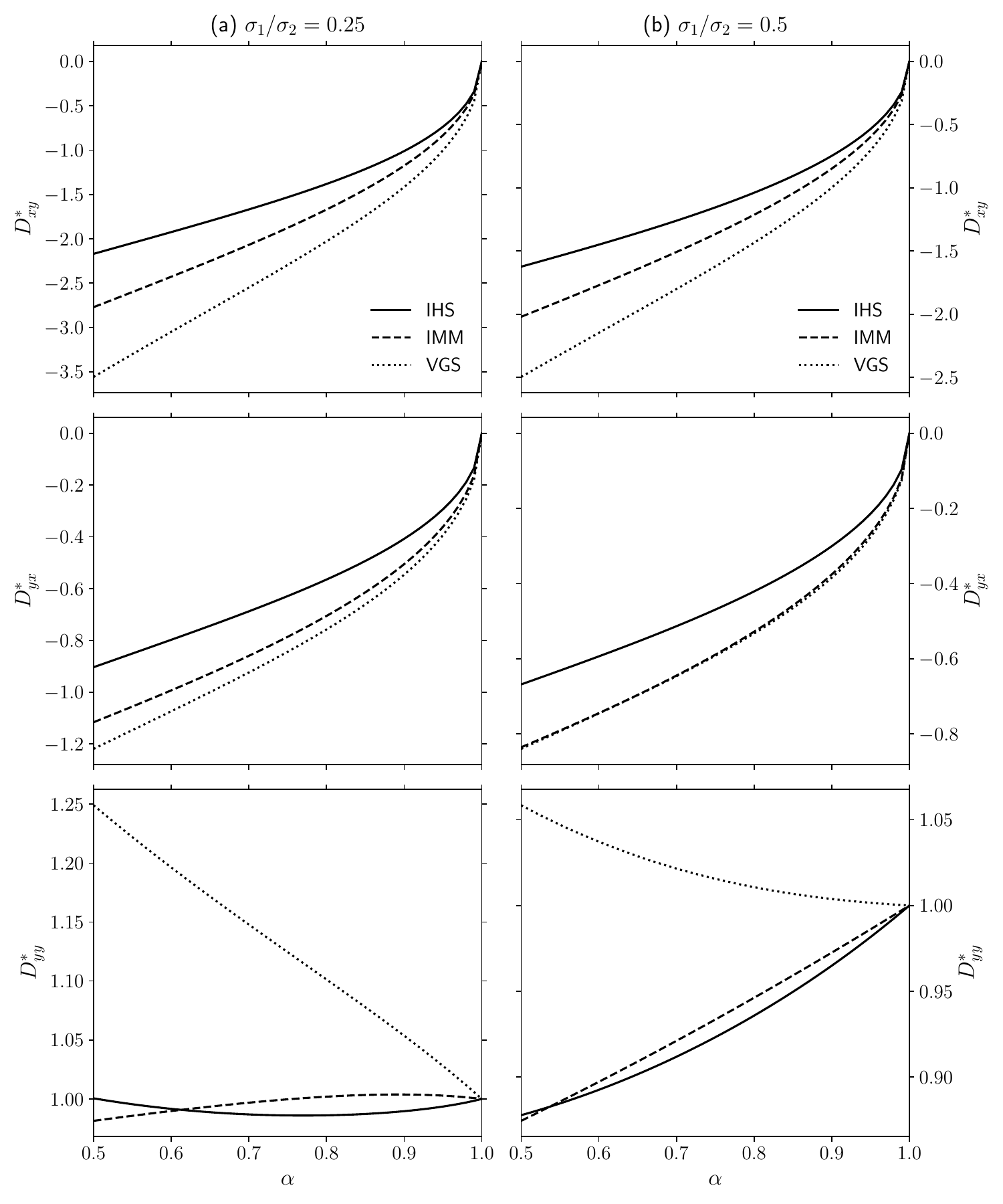}}
\end{tabular}
\end{center}
\caption{Panel (a): Plot of the (reduced) elements $D_{xy}^*$, $D_{yx}^*$, and $D_{yy}^*$ as functions of the (common) coefficient of restitution $\al$ for a three-dimensional system for $m_1/m_2=0.5$ and $\sigma_1/\sigma_2=0.25$. The solid lines correspond to the IHS results, the dashed lines refer to IMM results and the dotted lines are for the VGS results. Panel (b): The same as for the panel (a) for the case $m_1/m_2=0.5$ and $\sigma_1/\sigma_2=0.5$.
\label{compDij}}
\end{figure}

Now, we assume that the tracer and particles of the granular gas are mechanically different. The relevant elements of $D_{k\ell}^*$, $D_{p,k\ell}^*$, and $D_{T,k\ell}^*$ are plotted in figures \ref{fig3}, \ref{fig4}, and \ref{fig5}, respectively, as functions of $\al$ for the mixture $\sigma_1/\sigma_2=1$ and $m_1/m_2=0.5$. As occurs for the self-diffusion tensor, we observe that in general the influence of inelasticity on mass transport is quite significant regardless of the approximation used. Additionally, the anisotropy of the system (as measured by the differences $|T_{xx}-T_{yy}|$ and $|T_{zz}-T_{yy}|$) is much important in the shear flow $xy$-plane. In fact, while $T_{zz}-T_{yy}=0$ for IMM and kinetic model results, we observe that $T_{zz}\simeq T_{yy}$ for IHS. With respect to the comparison between IHS, IMM, and kinetic model, it is quite apparent that although the predictions of the kinetic model reproduce qualitatively well the results obtained for IHS, discrepancies between both approaches are found for strong dissipation. As occurs for the rheology, the agreement between IMM and IHS is better than the one found between IHS and the kinetic model. In fact, the good agreement obtained here between IHS and IMM can justify the use of IMM as a reliable model to unveil in a clean way the effect of inelasticity on transport in real granular flows.

To more clearly quantify the deviations between the results derived from IMM, VGS, and IHS, we define the following differences: $\Delta_{ij}^*=D_{ij}^{*}-D_{ij}^{*\text{IHS}}$, $\Delta_{p,ij}^*=D_{p,ij}^{*}-D_{p,ij}^{*\text{IHS}}$, and $\Delta_{T,ij}^*=D_{T,ij}^{*}-D_{T,ij}^{*\text{IHS}}$. Here, $D_{ij}^*$, $D_{p,ij}^*$, and $D_{T,ij}^*$ refer to the results obtained from IMM and the VGS kinetic model. Figures \ref{Deltaij}, \ref{Deltapij}, and \ref{DeltaTij} show the dependence of the above differences on $\al$ for IMM (panel (a)) and VGS (panel (b)). We have considered two different mixtures: $m_1/m_2=0.25$ and $\sigma_1/\sigma_2=0.5$ and $m_1/m_2=0.25$ and $\sigma_1/\sigma_2=1.5$. For the sake of concreteness, we have studied the elements $xy$, $yx$ and $yy$ of the above differences. As expected, we observe that the differences between IHS and IMM (and/or VGS) generally increase with inelasticity. These differences decrease as the diameter ratio $\sigma_1/\sigma_2$ increases (namely, as the tracer becomes larger than the particles of the granular gas). Depending on the element of $\Delta_{ij}^*$, $\Delta_{p,ij}^*$, and $\Delta_{T,ij}^*$ considered, the agreement between the IMM and IHS results is better or worse than those obtained from the VGS model. For example, at a diameter ratio of $\sigma_1/\sigma_2=1.5$ and a (common) coefficient of restitution of $\al=0.5$, the relative differences between the IMM and IHS results for the tracer diffusion coefficient elements $D_{xy}^*$, $D_{yx}^*$, and $D_{yy}^*$ are about 11\%, 32\%, and 2\%, respectively. The differences between the VGS and IHS results are about 15\%, 3\%, and 6\% for the elements $xy$, $yx$, and $yy$, respectively.

Finally, to assess the impact of particle size distribution on diffusion, we  consider a mixture with $m_1/m_2=0.5$ and two different diameter ratios. For clarity, figure \ref{compDij} shows only the elements of the tracer diffusion tensor, which is the most relevant quantity in a diffusion problem. As before, the IMM results compare better with the IHS results than with those obtained from the VGS model.

\section{Thermal diffusion segregation}
\label{sec7}

The knowledge of the diffusion transport coefficients allows us to apply the present theoretical results to the problem of segregation of tracer particles in a sheared granular gas. Segregation of dissimilar species in a granular mixture is likely one of the most relevant problems in granular flows not only from a fundamental point of view, but also from a more practical perspective. The problem was already studied in the context of IHS \cite{GV10}. Our objective here is to revisit the problem by employing the results derived for IMM and from the VGS kinetic model.

One of the most well-known examples of size segregation in vertically vibrated mixtures is the Brazil-nut effect (BNE), in which a relatively large particle attempts to move to the top of the sample against gravity \cite{RSPS87,KJN93,DRC93}. In addition, the reverse Brazil-nut effect (RBNE) has been observed in some experiments \cite{SM98,HQL01}. Several mechanisms  (void filling \cite{RSPS87}, convection \cite{KJN93}, arching \cite{DRC93}) have been proposed to explain the transition BNE/RBNE. Aside from these mechanisms, thermal diffusion becomes most relevant in strongly shaken or sheared granular systems, as the motion of grains in these systems is similar to the chaotic motion of atoms or molecules in an ordinary gas.

Thermal diffusion (or thermophoresis) is one of the most extensively studied phenomena in ordinary gases and liquids \cite{GI52,KCL87}. Thermal diffusion refers to the transport of matter due to the presence of a thermal gradient. This motion generates concentration gradients in the system, resulting in diffusion in the mixture. After a transient regime, the system reaches a steady state in which the separation effect arising from thermal diffusion is balanced by pure diffusion. The existence of these competing mechanisms results in the segregation (or partial separation) of the mixture's different species.

Although the study of segregation by thermal diffusion in granular mixtures using kinetic theory tools (see for example, Refs.\ \cite{SGNT06,G19}) in the Navier--Stokes domain is well-established, much less is known about sheared granular mixtures. One reason is that a complete description of the segregation problem in sheared systems requires introducing a thermal diffusion tensor to characterize segregation in different directions. This is a consequence of cross-effects appearing in mass transport when the system is under USF. Here, as in  \cite{GV10}, for the sake of simplicity we consider a situation where the temperature gradient is perpendicular to the shear flow plane. Thus, for a three-dimensional system, $\partial_x T=\partial_y T=0$ but $\partial_z T\neq 0$. Additionally, since we are interested in a situation where the hydrodynamic equations \eqref{4.2}--\eqref{4.4} admit a steady solution, we also assume that $\delta \mathbf{U}=\mathbf{0}$. Under these conditions, the amount of segregation parallel to the $z$-axis can be measured by the thermal diffusion factor $\Lambda_z$ defined as
\beq
\label{7.1}
-\Lambda_z \frac{\partial \ln T}{\partial z}=\frac{\partial \ln x_1}{\partial z}.
\eeq
Let us assume that the thermal gradient is directed downwards ($\partial_z T<0$). In this case, when $\Lambda_z>0$, according to equation \eqref{7.1} then $\partial_z \ln x_1>0$. This means that in this case the tracer particles tend to accumulate near the cold wall. On the other hand, when $\Lambda_z<0$, the tracer particles tend to move towards the hot wall since $\partial_z \ln x_1<0$. In other words, the signature of $\Lambda_z$ provides a segregation criterion for the tracer particles immersed in a sheared granular gas.

Let us write $\Lambda_z$ in terms of the pressure tensors $P_{zz}$ and $P_{1,zz}$ as well as the diffusion coefficients $D_{zz}$, $D_{p,zz}$ and $D_{T,zz}$. First,  when only gradients along the $z$-axis exist, the momentum balance equation \eqref{4.3} yields
\beq
\label{7.2}
\frac{\partial P_{zz}}{\partial z}=0.
\eeq
Since $P_{zz}=pP_{zz}^*(a^*)$, then
\beq
\label{7.3}
\frac{\partial P_{zz}}{\partial z}=\frac{\partial p}{\partial z}\left(1-a^*\partial_{a^*}\right)P_{zz}^*+
\frac{p}{2T}\frac{\partial T}{\partial z}a^*(\partial_{a^*}P_{zz}^*),
\eeq
where $\partial_{a^*}P_{zz}^*=\partial_{a^*}P_{yy}^*\equiv \Upsilon_{yy}$ is given by Eq.\ \eqref{b4}. Using equations\ \eqref{7.2} and \eqref{7.3}, one gets a relationship between the quantities $\partial_z p$ and $\partial_z T$:
\beq
\label{7.4}
\frac{\partial \ln p}{\partial z}=-\frac{1}{2}\frac{a^*(\partial_{a^*}P_{zz}^*)}{P_{zz}^*-a^*(\partial_{a^*}P_{zz}^*)}\frac{\partial \ln T}{\partial z}.
\eeq
In addition, according to the balance equation \eqref{4.2}, in the steady state with $\delta \mathbf{U}=\mathbf{0}$ then $j_{1,z}^{(1)}=0$. The constitutive equation for $j_{1,z}^{(1)}$ is
\begin{equation}
\label{7.5}
j_{1,z}^{(1)}=-m_1 D_{zz}\partial_z x_1-\frac{m_2}{T}D_{p,zz}\partial_z p-\frac{\rho}{T}D_{T,zz}\partial_z T.
\end{equation}
The condition $j_{1,z}^{(1)}=0$ leads to
\beq
\label{7.6}
\frac{\partial \ln x_1}{\partial z}=-\frac{\rho}{m_1 x_1}\left(\frac{D_{p,zz}}{D_{zz}}\frac{\partial \ln p}{\partial z}
+\frac{D_{T,zz}}{D_{zz}}\frac{\partial \ln T}{\partial z}\right)=
-\Bigg[(1-\mu)\frac{D_{p,zz}^*}{D_{zz}^*}\frac{\partial \ln p}{\partial z}+\Omega^{\text{el}*}
\frac{D_{T,zz}^*}{D_{zz}^*}\frac{\partial \ln T}{\partial z}\Bigg].
\eeq
Here, we recall that $\mu=m_1/m_2$ is the mass ratio and $\Omega^{\text{el}*}= \Omega^{\text{el}}/\nu$, where $\Omega^{\text{el}}$ is defined by equation \eqref{4.25} and the value of $\nu$ depends on the approach followed.

From equations\ \eqref{7.4} and \eqref{7.6} one easily gets the expression of the thermal diffusion factor $\Lambda_z$ as
\beq
\label{7.8}
\Lambda_z=\frac{\Omega^{\text{el}*}D_{T,zz}^*-\frac{1}{2}(1-\mu)a^*\Upsilon_{yy}\left(P_{zz}^*-a^*\Upsilon_{yy}\right)^{-1}
D_{p,zz}^*}{D_{zz}^*}.
\eeq
The condition $\Lambda_z=0$ gives the criterion for the upwards/downwards segregation transition.
\begin{figure}[t]
\begin{center}
\begin{tabular}{lr}
\resizebox{9.5cm}{!}{\includegraphics{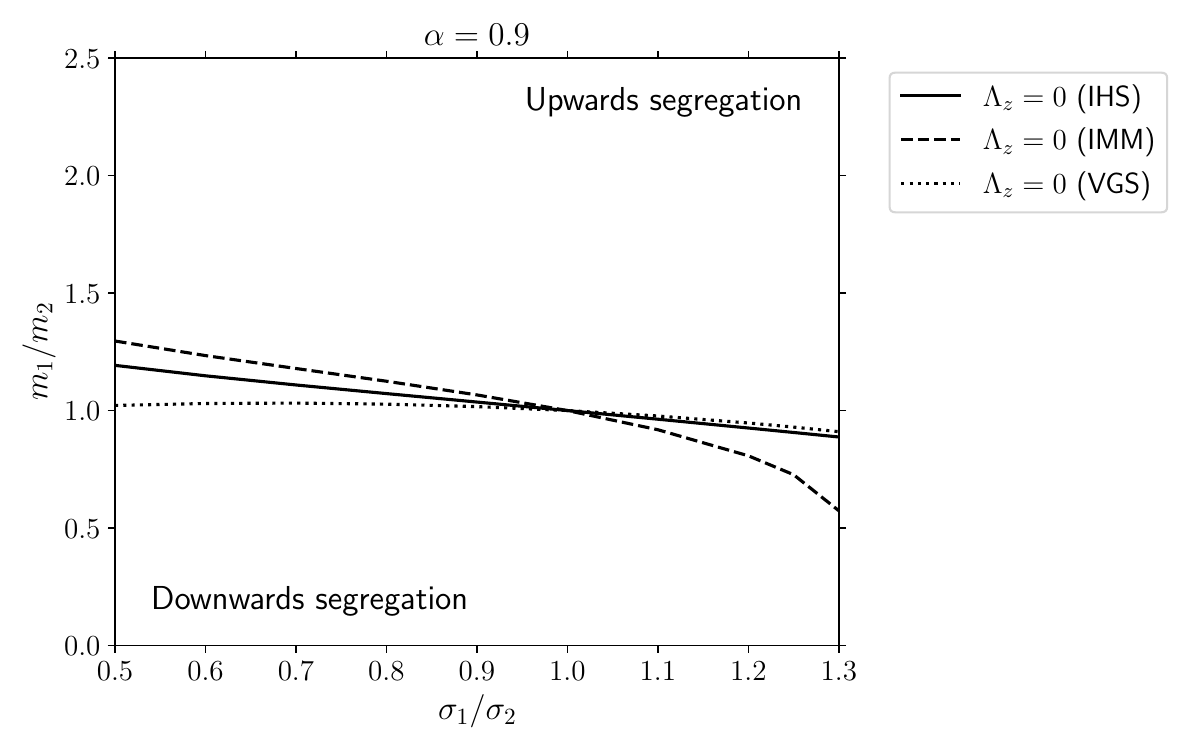}}
\end{tabular}
\begin{tabular}{lr}
\resizebox{9.5cm}{!}{\includegraphics{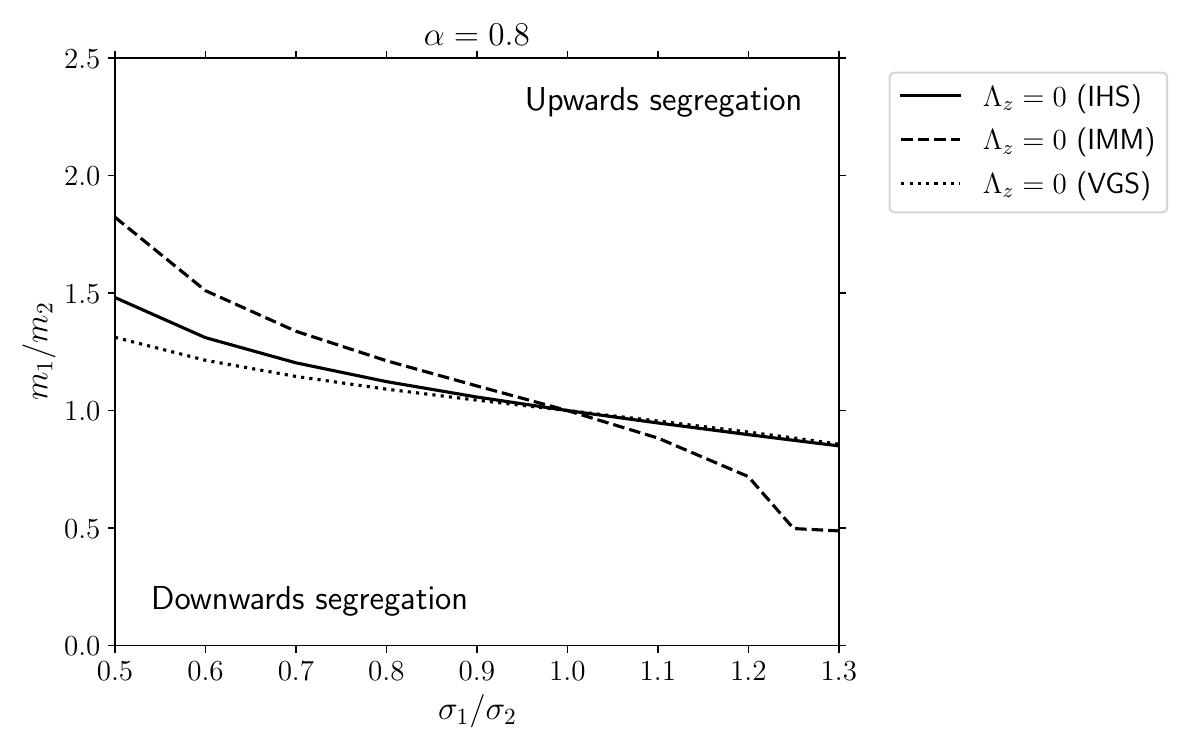}}
\end{tabular}
\end{center}
\caption{(a) Phase diagram for segregation in the $\left\{\sigma_1/\sigma_2, m_1/m_2\right\}$ plane for a three-dimensional system ($d=3$) with $\al=0.9$. (b) Phase diagram for segregation in the $\left\{\sigma_1/\sigma_2, m_1/m_2\right\}$ plane for a three-dimensional system ($d=3$) with $\al=0.8$.
\label{fig6}}
\end{figure}
In accordance to Eq.\ \eqref{4.24} the diffusion coefficient $D_{zz}^*$ is positive. As a consequence, the marginal segregation curve ($\Lambda_z=0$) is obtained from the condition
\beq
\label{7.9}
\Omega^{\text{el}*}D_{T,zz}^*-\frac{1}{2}(1-\mu)a^*\Upsilon_{yy}\left(P_{zz}^*-a^*\Upsilon_{yy}\right)^{-1}
D_{p,zz}^*=0.
\eeq

According to equation \eqref{7.9}, there are clearly several competing mechanisms in thermal diffusion segregation. First, since the granular gas is in a far-from-equilibrium state, $P_{zz}^*$ differs from 1 and the derivative $\Upsilon_{yy}$ differs from 0. Second, the forms of the diagonal elements $D_{T,zz}^*$ and $D_{p,zz}^*$ differ from their corresponding elastic forms. These deviations essentially arise from the fact that the reference shear flow state of the tracer particles being completely different from that of the granular gas particles. Notably, energy equipartition breaks down since the tracer temperature $T_1$ differs from the granular temperature $T_2\simeq T$. This effect has a significant impact on granular segregation.

\subsection{Some limiting cases}

Before considering the general case, it is convenient to consider some simple cases. First, for elastic collisions ($\al_{22}=\al_{12}=1$), $D_{T,zz}^*=0$ and equation \eqref{3.6} for IMM or equation \eqref{5.9} for the VGS kinetic model yields $a^*=0$. Thus, the condition \eqref{7.9} is trivially satisfied for any value of the mass and diameter ratios. This means that no segregation occurs in this limiting case as expected.

Another interesting case refers to a granular mixture constituted by mechanically equivalent species. In this situation, although the collisions are inelastic, $D_{p,k\ell}^*=D_{T,k\ell}^*=0$. This means that
$\Lambda_z=0$ for any value of the coefficient of restitution. This is the expected result since both species are indistinguishable.

\subsection{General case}

For inelastic collisions, the zero contour of $\Lambda_z$ exhibits a complex nonlinear dependence on the parameter space of the system. Thus, as did in section \ref{sec6}, for the sake of simplicity we take a three-dimensional granular gas ($d=3$) in the case of a common coefficient of restitution ($\al_{22}=\al_{12}\equiv \al$). The marginal segregation curve $\Lambda_z=0$ separates regions of $\Lambda_z>0$ (upwards segregation) and $\Lambda_z<0$ (downwards segregation). At a fixed value of $\al$, the points lying on the zero contour correspond to values of the diameter and mass ratios for which the intruder does not segregate in a sheared granular gas.

As an illustration, figure \ref{fig6} shows the phase diagram in the $\left(\sigma_1/\sigma_2, m_1/m_2\right)$-plane for two different values of the (common) coefficient of restitution $\al$. We compare the theoretical predictions for the marginal segregation curve $\Lambda_z=0$ obtained previously for IHS in  \cite{GV10} with those derived here for IMM and from the VGS kinetic model. As previously mentioned, all curves pass through the point $(1,1)$ because it corresponds to the limiting case of mechanically equivalent particles. The three approaches show that, for $\sigma_1<\sigma_2$, the main effect of inelasticity (or equivalently, the reduced shear rate $a^*$) is to enlarge the size of the downwards segregation region. The opposite occurs when the tracer particles are larger than the particles of the granular gas. In general, we see that the tracer particles tend to move toward hotter regions since upwards segregation occupies most of the system's parameter space. Additionally, the results obtained here for sheared granular gases differ qualitatively from the segregation results obtained for systems with no shear (i.e., vibrated dense systems). In the latter case, segregation tends to predominantly be of the downward type as the size of the tracer particles increases. This is in fact the opposite behavior observed here for strongly sheared granular gases.

Regarding the comparison of the three approaches of figure \ref{fig6}, the results derived from IMM qualitatively agree well with the IHS results. Surprisingly, however, the segregation results obtained from the VGS model agree better with the IHS results than with the IMM results. This good agreement is especially noticeable when $\sigma_1>\sigma_2$.

\section{Discussion}
\label{sec8}

In this paper, we have analyzed the diffusion of tracer particles immersed in a sheared granular gas. Under these conditions, the mass flux $\mathbf{j}_1^{(1)}$ is defined in terms of the tracer diffusion tensor $D_{k\ell}$ (which couples $\mathbf{j}_1^{(1)}$ with the concentration gradient $\nabla x_1$), the pressure diffusion tensor $D_{p,k\ell}$ (which couples $\mathbf{j}_1^{(1)}$ with the pressure gradient $\nabla p$), and the thermal diffusion tensor $D_{T,k\ell}$ (which couples $\mathbf{j}_1^{(1)}$ with the temperature gradient $\nabla T$). These tensorial quantities were evaluated years ago in the context of the Boltzmann equation for IHS \cite{G02,G07}. However, due to the intricate mathematical structure of the Boltzmann collision operator for IHS, the results obtained in Refs.\ \cite{G02,G07} involve several (uncontrolled) approximations at different stages of the derivation. Here, we revisit this problem by considering two different, complementary approaches that allow us to achieve exact results. First, we maintain the structure of the Boltzmann collision operators but consider a different interaction model: the so-called IMM, in which the collision rate of colliding spheres is independent of their relative velocity. This simplification enables us to obtain exact expressions for the rheological properties of the system (granular gas plus tracer particles), as well as the diffusion tensors. As a second approach, we keep the IHS interaction model but replace the true Boltzmann collision operators with simpler mathematical terms that retain their relevant physical properties. In this context, we consider the VGS kinetic model \cite{VGS07} proposed years ago for granular mixtures.

As in the paper \cite{G07} for IHS, the diffusion tensors are obtained by solving the Boltzmann equation (or the VGS model) for tracer particles using a generalization of the Chapman--Enskog method \cite{CC70} for far-from-equilibrium states. Since the granular gas is subjected to a strong shear rate, non Newtonian effects are relevant for finite inelasticity. Thus, the reference state (the zeroth-order distribution $f_1^{(0)}$) in the perturbation method is the shear flow distribution, not the local equilibrium distribution. Additionally, since collisional cooling cannot compensate for viscous heating locally, $f_1^{(0)}$ is in general a time-dependent distribution even when the gas is slightly perturbed from the USF. Once the linear integral equations verifying the diffusion tensors are obtained, we restrict to steady state conditions and so, the reduced shear rate $a^*$ is coupled to the coefficient of restitution $\al_{22}$ which characterizes the inelasticity of grain-grain collisions.  The consideration of the steady state allows us to achieve analytical, exact expressions for $D_{k\ell}$, $D_{p,k\ell}$, and $D_{T,k\ell}$. These tensors depend nonlinearly on the diameter ratio, $\sigma_1/\sigma_2$, the mass ratio, $m_1/m_2$, and the coefficients of restitution $\al_{22}$ and $\al_{12}$ (which characterize the inelasticity of tracer-grain collisions).

The results obtained from the IMM and VGS models for the diffusion tensors show that the non-zero elements of these tensors depend intricately on the coefficients of restitution $\al_{22}$ and $\al_{12}$, as well as on the masses and diameters of the mixture. According to equations  \eqref{4.19}--\eqref{4.21}, we can conclude that the deviations of the coefficients $(D_{ij}, D_{p,ij}, D_{T,ij})$ from their elastic values are due to three main reasons. First,
the shear field alters the conventional collision frequency of the elastic diffusion coefficient $\Omega^{\text{el}}$ (defined by equation \eqref{4.25}) by the tensor $a_{ij}+\Omega_{ij}$, where $\Omega_{ij}=\Omega \delta_{ij}$ for IMM and $\Omega_{ij}=\Omega' \delta_{ij}$ for the VGS kinetic model. Second, since the tracer and granular gas particles are generally mechanically different, their corresponding rheological properties differ as well. This feature gives rise to the presence of the tracer pressure tensor $P_{1,k\ell}^{(0)}$ instead of $P_{2,k\ell}^{(0)}$ in equation \eqref{4.19} for $D_{ij}$ and the
terms of the form $(m_1/m_2)P_{1,k\ell}^{(0)}-P_{2,k\ell}^{(0)}$ in equations \eqref{4.20} and \eqref{4.21}, which define the tensors $D_{p,ij}$ and $D_{T,ij}$, respectively. Third, the USF state is inherently non-Newtonian, so there is a coupling between the coefficients $D_{p,ij}$ and $D_{T,ij}$. Since the (reduced) shear rate $a^*$ (which measures the departure of the system from equilibrium) is coupled to the coefficient of restitution $\al_{22}$ via equation \eqref{3.6.1}, the above three effects disappear when the collisions are elastic.

A comparison of the exact theoretical results of the IMM and VGS models with the approximate IHS results generally shows good qualitative agreement. At a more quantitative level, we observe excellent agreement in some cases (e.g., rheological properties) and reasonably good agreement in others (e.g., diffusion tensors), especially in the case of IMM. It is quite apparent that to confirm the reliability of the predictions offered by IMM and VGS model for the tracer diffusion coefficients one should compare them with computer simulation results for IHS. At the level of rheology, the results reported here (see figure \ref{fig1}) and in \cite{G03} for a binary mixture of IMM under USF clearly demonstrate the accuracy of this interaction model to capture the dependence of the pressure tensors on the coefficients of restitution.

To complement this analysis, we have also addressed the sensitivity of the tracer diffusion transport coefficients ($D_{ij}^*, D_{T,ij}^*, D_{p,ij}^*$) to the system parameters
$$
\{\alpha_{12}, \alpha_{22}, m_1/m_2, \sigma_1/\sigma_2,d\}.
$$
Since the space parameter of the system (granular gas plus tracer particles) is relatively high, we have considered a three-dimensional ($d=3$) system with a \textit{common} coefficient of restitution $\al_{22}=\al_{12}\equiv \al$, to gain insight into the general problem. In this case, the diffusion coefficients depend on three independent parameters: the mass and diameter ratios as well as the (common) coefficient of restitution.
Although the highly non-linear dependence of these transport coefficients on the aforementioned quantities makes it challenging to isolate the influence of each parameter, our results identify the dominant trends. As for the (common) coefficient of restitution $\alpha$, figures \ref{fig3}--\ref{fig5} show the correspondence between a higher degree of inelasticity (stronger anisotropy) and the exacerbation of the quantitative differences among the interaction models considered.
Additionally, we find that departures of the diffusion  coefficients from their elastic functional forms are generally substantial, even when the level of dissipation is only moderate.
Regarding the influence of the diameter ratio for fixed values of $\alpha$ and mass ratio $m_1/m_2$, figures \ref{Deltaij}--\ref{DeltaTij} show that the discrepancies between IHS and IMM (and/or VGS) tend to decrease as the tracer particles become larger than those of the granular gas (i.e. when $\sigma_1/\sigma_2$ increases). While no universal hierarchy can be established concerning the relative accuracy of the exact VGS and IMM results compared to the approximated IHS results, the latter generally exhibits a closer agreement with IHS in the majority of the cases studied in the paper.

As a nice application of the results exposed in this paper, the segregation of tracer particles in a sheared granular gas has been analyzed. The relative motion of the tracers with respect to the particles of the gas is caused by the presence of a temperature gradient. Here, for the sake of simplicity, we have assumed that the thermal gradient $\partial_z T$ is perpendicular to the shear flow $xy$-plane. Under these conditions, the amount of segregation in the $z$-direction is measured by the thermal diffusion factor $\Lambda_z$, defined in Eq.\ \eqref{7.1}. The condition of zero thermal diffusion ($\Lambda_z=0$) gives the segregation criterion for the transition from upwards segregation (regions where $\Lambda_z>0$) to downwards segregation (regions where $\Lambda_z<0$). A comparison with previous results obtained for IHS \cite{GV10} (see figure \ref{fig6}) shows reasonable agreement in general, especially for the VGS model when the tracer particles are larger than the granular gas particles.

Beyond the comparison among the results offered by the different interaction models for segregation, the sensitivity of the segregation criterion \eqref{7.9} to the mechanical properties is also explicitly captured in the phase diagrams derived in Section \ref{sec7}. In this regard, both the mass $m_1/m_2$ and diameter $\sigma_1/\sigma_2$ ratios act as the critical factors governing the sign of the thermal diffusion factor $\Lambda$ at fixed $\alpha$. For $\sigma_1<\sigma_2$, a higher degree of inelasticity tends to enlarge the size of the downwards segregation region, while the opposite occurs when $\sigma_1 > \sigma_2$. Concurrently, a lower value of the (common) coefficient of restitution $\alpha$ is usually associated with an expansion of the region in which tracer particles with a mass ratio $m_1/m_2 > 1$ segregate downward, as shown by figure \ref{fig6}.

As said before, a more quantitative comparison between the theoretical results derived here for IMM and VGS with those obtained by molecular dynamics or Monte Carlo (DSMC) simulations would allow us to gauge the reliability of the present results. However, we are not aware of any work in the granular literature where the diffusion coefficients of tracer particles immersed in a strongly sheared \textit{dilute} gas have been measured. Thus,
the lack of simulation data for the diffusion coefficients in the low-density regime prevents a comparison between theory and simulation. However, as mentioned throughout the paper,  previous papers \cite{C97,ALJR21} have analyzed the $\al$-dependence of the self-diffusion tensor for very dense systems. The density range examined in these papers precludes a quantitative comparison. We hope that this paper will encourage simulators to perform simulations under shear conditions similar to those considered here to confirm the theoretical results displayed in this paper. In particular, one could perform Monte Carlo simulations in a sheared granular mixture by following Campbell's strategy \cite{C97} for computing the self-diffusion tensor using particle tracking and velocity correlations.

\vspace{3.5mm}
\subsection{Conclusion}

In conclusion, the results found here give evidence of the accuracy of both IMM and the VGS kinetic model for studying far-from-equilibrium situations in granular flows, where using the original Boltzmann equation for IHS turns out to be quite intricate. Additionally, using a kinetic model instead of the Boltzmann equation for IHS and/or IMM allows us to obtain the explicit forms of the velocity distribution functions. This is likely one of the main advantages of starting from a kinetic model rather than the true Boltzmann kinetic equation. In particular, as a future project, we plan to use the VGS kinetic model to determine the shear-rate dependent tracer diffusion  coefficients in a sheared granular suspension (namely, a granular gas immersed in a molecular gas).

\acknowledgments

V.G. acknowledges financial support from Grant No. PID2024-156352NB-I00
funded by \linebreak MCIU/AEI/10.13039/501100011033/FEDER, UE and from Grant No.
GR24022 funded by Junta de Extremadura (Spain) and by European Regional
Development Fund (ERDF) ``A way of making Europe''.

\appendix
\setcounter{section}{0}
\section{Linear stability analysis of the steady state solutions to USF}
\label{appA}

In this Appendix, we want to see whether the steady state solutions \eqref{3.5}--\eqref{3.12} to the pressure tensors are (linearly) stable. We consider first the time evolution equation for the pressure tensor of the excess granular gas. The three relevant independent equations for $P_{2,k\ell}$ are given by
\beq
\label{a1}
\partial_t p+\zeta_2 p+\frac{2a}{d}P_{2,xy}=0,\quad \partial_t P_{2,xy}+\nu_\eta P_{2,xy}+a P_{2,yy}=0,
\eeq
\beq
\label{a1bis}
\partial_t P_{2,yy}+\nu_\eta P_{2,yy}-\left(\nu_\eta-\zeta_2\right) P_{2,yy}=0,
\eeq
where $\nu_\eta=\nu_{\text{M},22}\nu_\eta^*$ and $p\simeq p_2=n_2 T_2$. In terms of the dimensionless quantities, $P_{2,k\ell}^*(t)=P_{2,k\ell}(t)/n_2 T(t)$, $a^*(t)=a/\nu_{\text{M},22}(t)$ and
\beq
\label{a2}
\tau(t)=\int_0^{t} \mathrm{d}t'\; \nu_{\text{M},22}(t'),
\eeq
equations\ \eqref{a1} and \eqref{a1bis} become
\beq
\label{a3}
2\partial_\tau \ln a^*=\zeta_2^*+\frac{2}{d}a^* P_{2,xy}^*,
\eeq
\beq
\label{a4}
\partial_\tau P_{2,xy}^*=-a^*P_{2,yy}^*-P_{2,xy}^*\Big(\nu_\eta^*-\zeta_2^*-\frac{2}{d}a^* P_{2,xy}^*\Big),
\eeq
\beq
\label{a5}
\partial_\tau P_{2,yy}^*=-P_{2,yy}^*\Big(\nu_\eta^*-\zeta_2^*-\frac{2}{d}a^* P_{2,xy}^*\Big)+\nu_\eta^*-\zeta_2^*,
\eeq
The variable $\tau$ is the dimensionless time measured as the average collision number. A steady solution of equations\ \eqref{a3}--\eqref{a5} is given by equations\ \eqref{3.5} and \eqref{3.6}. To carry out a linear stability analysis of these steady solutions, we look for solutions to the set \eqref{a3}--\eqref{a5} given by
\beq
\label{a6}
a^*(\tau)=a_s^*+\delta a^*(\tau), \quad P_{2,xy}^*(\tau)=P_{2xy,s}^*+\delta P_{2,xy}^*(\tau), \quad P_{2,yy}^*(\tau)=P_{2yy,s}^*+\delta P_{2,yy}^*(\tau),
\eeq
where the subscript $s$ means that the quantities are evaluated in the steady state. Substituting the identities \eqref{a6} into equations\ \eqref{a3}--\eqref{a5} and neglecting nonlinear terms in the perturbations, one gets
\beq
\label{a7}
\partial_\tau
\left(
\begin{array}{c}
\delta a_s^*\\
\delta P_{2,xy}^*\\
a_s^*\delta P_{2,yy}^*
\end{array}
\right)=-\mathsf{L}\cdot \left(
\begin{array}{c}
\delta a^*\\
\delta P_{2,xy}^*\\
a_s^*\delta P_{2,yy}^*
\end{array}
\right),
\eeq
where $\mathsf{L}$ is the square matrix
\begin{equation}
\mathsf{L}=\left(
\begin{array}{ccc}
\frac{\zeta_2^*}{2} & -\frac{\zeta_2^*\nu_\eta^{*2}}{2(\nu_\eta^*-\zeta_2^*)} & 0 \\
\left(\frac{\nu_\eta^*-\zeta_2^*}{\nu_\eta^*}\right)^2 & \nu_\eta^*+\zeta_2^*& 1\\
\frac{(\nu_\eta^*-\zeta_2^*)\zeta_2^*}{\nu_\eta^*} & -\zeta_2^*\nu_\eta^* & \nu_\eta^*
\end{array}
\right).  \label{a8}
\end{equation}

The time evolution of the deviations from the steady solution is governed by the three eigenvalues of the matrix $\mathsf{L}$. If the real parts of those eigenvalues are positive the steady solution is linearly stable, while is unstable otherwise. The eigenvalues are determined from the solution of the secular equation
\beq
\label{a9}
\det \left(L_{k\ell}-\ell \delta_{k\ell}\right)=0.
\eeq
The solution to Eq.\ \eqref{a9} leads to a real eigenvalue $\ell_1$ and a pair of complex conjugate eigenvalues $\ell_2$ and $\ell_3$. As for IHS, the results show that $\text{Re}(\lambda_i)>0$ ($i=1,2,3$) for any value of $\al_{22}$. This means that the steady USF solution for the excess granular is linearly stable, and the characteristic relaxation time (measured by the number of collisions) is $\ell_1^{-1}$. As an illustration, figure \ref{appAfig1} shows the dependence of $\ell_1$ and of the real part of $\ell_{2,3}$ for a three-dimensional granular gas ($d=3$). Note that $\ell_1\to 0$ in the elastic limit $\al_{22}\to 1$. This is a consequence that for elastic collisions $a^*\to 0$.

\begin{figure}[t]
\begin{center}
\begin{tabular}{lr}
\resizebox{10.0cm}{!}{\includegraphics{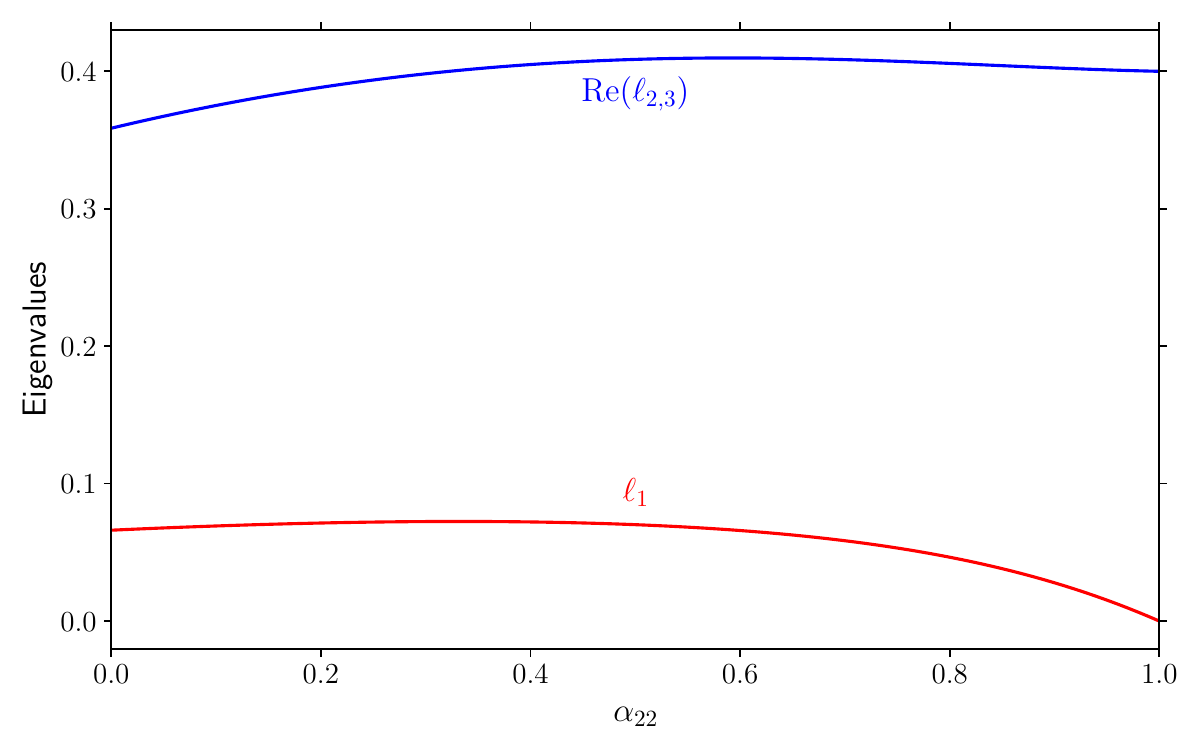}}
\end{tabular}
\end{center}
\caption{Plot of the eigenvalue $\ell_1$ and of the real part of $\ell_{2,3}$ as a function of the coefficient of restitution $\al_{22}$ for a three-dimensional system.
\label{appAfig1}}
\end{figure}

We consider now the (linear) stability of the steady solution to the set of time evolution equations associated with $\gamma(t)$, $P_{1,xy}(t)$ and $P_{1,yy}(t)$. Since we have previously shown that the perturbations $(\delta a^*, \delta P_{2,xy}^*, \delta P_{2,yy}^*)$ tend to zero for sufficiently long times, we assume hence that $a^*\equiv \text{const.}$, $P_{2,xy}^*\equiv \text{const.}$ and $P_{2,yy}^*\equiv \text{const.}$ in the evolution equations of $\gamma(t)$, $P_{1,xy}(t)$ and $P_{1,yy}(t)$. In terms of the variable $\tau$, the set of equations for $\gamma$, $P_{1,xy}$ and $P_{1,yy}$ is
\beq
\label{a10}
\partial_\tau \gamma=-\frac{2}{d}a^* P_{1,xy}^*-\zeta_1^* \gamma,
\eeq
\beq
\label{a11}
\partial_\tau P_{1,yy}^*=Y+ X_0 P_{1,yy}^*+X P_{2,yy}^*,
\eeq
\beq
\label{a12}
\partial_\tau P_{1,xy}^*+a^* P_{1,yy}^*= X_0 P_{1,xy}^*+X P_{2,yy}^*,
\eeq
where the quantities $Y$, $X_0$ and $X$ are defined by equations\ \eqref{3.8} and \eqref{3.9}, respectively.  As in the case of the excess granular gas, we want to solve the set of equations\ \eqref{a10}--\eqref{a12} by assuming small deviations from the steady state solution. Thus, we write
\beq
\label{a13}
\gamma(\tau)=\gamma_s+\delta \gamma (\tau), \quad P_{1,yy}^*(\tau)=P_{1yy,s}^*+\delta P_{1,yy}^*(\tau), \quad P_{1,xy}^*(\tau)=P_{1xy,s}^*+\delta P_{1,xy}^*(\tau).
\eeq
In the linear order in the perturbations,
\beq
\label{a14}
\zeta_1^*=\zeta_{1s}^*+\overline{\zeta}_1 \delta \gamma, \quad Y=Y_s+\overline{Y}\delta\gamma,\quad X_0=X_{0s}+\overline{X}_0\delta\gamma, \quad X=X_s+\overline{X}\delta\gamma,
\eeq
where $\zeta_{1s}^*$, $Y_s$, $X_{0s}$, and $X_s$ refer to the values of these quantities in the steady state and
\begin{align}
\label{a15}
\overline{\zeta}_1=
\frac{\sqrt{2}}{d}\left(\frac{\sigma_{12}}{\sigma_2}\right)^{d-1}\mu_{21}(1+\al_{12})\Bigg\{
\frac{(1+\theta_s^{-1})^{-1/2}}{2\mu}\left[1-\frac{\mu_{21}}{2}(1+\al_{12})(1+\theta_s)\right]+\frac{\mu_{21}^2}
{2\mu_{12}}\theta_s^2(1+\theta_s^{-1})^{1/2}(1+\al_{12})
\Bigg\},
\end{align}

\begin{figure}[t]
\begin{center}
\begin{tabular}{lr}
\resizebox{10.0cm}{!}{\includegraphics{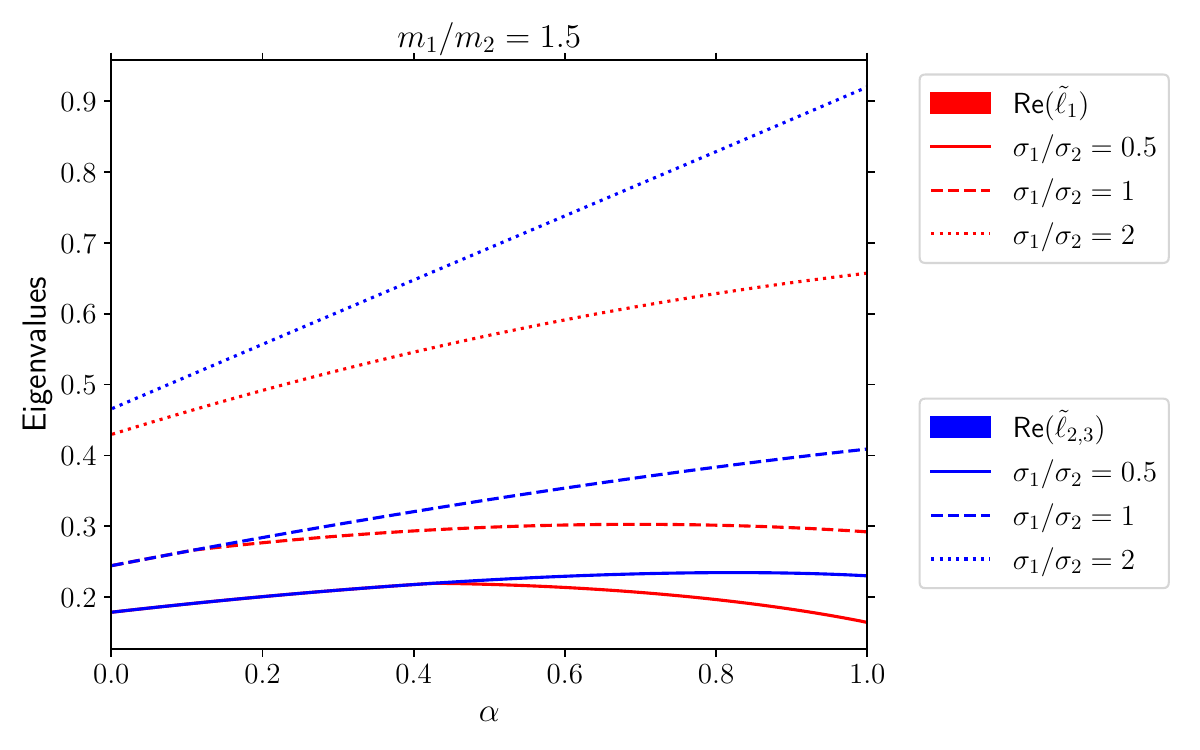}}
\end{tabular}
\end{center}
\caption{Plot of the real parts of the eigenvalues $\widetilde{\ell}_i$ ($i=1,2,3$) as functions of the (common) coefficient of restitution $\al_{22}=\al_{12}\equiv \al$ for the mass ratio $m_1/m_2=1.5$ and different values of the diameter ratio $\sigma_1/\sigma_2$.
\label{appAfig3}}
\end{figure}

\beq
\label{a16}
\overline{Y}=\frac{3}{2\sqrt{2}}\left(\frac{\sigma_{12}}{\sigma_2}\right)^{d-1}\frac{\mu_{21}^2}{(d+2)}(1+\al_{12})^2
(1+\theta_s^{-1})^{1/2},
\eeq
\beq
\label{a17}
\overline{X}_0=-\frac{1}{\sqrt{2}d(d+2)}\left(\frac{\sigma_{12}}{\sigma_2}\right)^{d-1}\frac{\mu_{21}^2}{\mu_{12}}
(1+\theta_s^{-1})^{-1/2}(1+\al_{12})\left[d+2-\mu_{21}(1+\al_{12})\right],
\eeq
\beq
\label{a18}
\overline{X}=\frac{1}{\sqrt{2}d(d+2)}\left(\frac{\sigma_{12}}{\sigma_2}\right)^{d-1}\mu_{21}^2
(1+\theta_s^{-1})^{-1/2}(1+\al_{12})^2.
\eeq
Substitution of equations\ \eqref{a13} and \eqref{a14} into equations\ \eqref{a10}--\eqref{a12} and neglecting nonlinear terms in the perturbations, after some algebra one gets the set of linear differential equations:
\beq
\label{a19}
\partial_\tau
\left(
\begin{array}{c}
\delta \gamma\\
\delta P_{1,yy}^*\\
\delta P_{1,xy}^*
\end{array}
\right)=-\widetilde{\mathsf{L}}\cdot \left(
\begin{array}{c}
\delta \gamma\\
\delta P_{1,yy}^*\\
\delta P_{1,xy}^*
\end{array}
\right),
\eeq
where $\widetilde{\mathsf{L}}$ is the square matrix
\begin{equation}
\widetilde{\mathsf{L}}=\left(
\begin{array}{ccc}
\zeta_{1s}^*+\gamma_s \overline{\zeta}_1 & 0& \frac{2}{d}a_s^* \\
-\left(\overline{Y}+\overline{X}_0P_{1,yy,s}^*+\overline{X}P_{yy,s}^*\right) & -X_{0s}& 0\\
-\left(\overline{X}_0P_{1,xy,s}^*+\overline{X}P_{yy,s}^*\right) & a_s^* & -X_{0s}
\end{array}
\right).  \label{a20}
\end{equation}
The eigenvalues of the matrix $\widetilde{\mathsf{L}}$ are the roots of the secular equation
\beq
\label{a21}
\det \left(\widetilde{L}_{k\ell}-\widetilde{\ell} \delta_{k\ell}\right)=0.
\eeq
If the real parts of the eigenvalues $\widetilde{\ell}$ are always positive for any value of the set $(\al_{22},\al_{12},m_1/m_2, \sigma_1/\sigma_2)$ then the steady USF solution for the pressure tensor $P_{1,ij}^*$ is linearly stable.

A systematic analysis of the dependence of $\text{Re}( \widetilde{\ell}_i)$ ($i=1,2,3$) on the parameter space shows that the real parts of the eigenvalues are always positive. As an illustration, figures \ref{appAfig2} and \ref{appAfig3} show the $\al$-dependence of $\text{Re}(\widetilde{\ell}_i)$; being quite apparent that their real parts are positive.

\begin{figure}[t]
\begin{center}
\begin{tabular}{lr}
\resizebox{10.0cm}{!}{\includegraphics{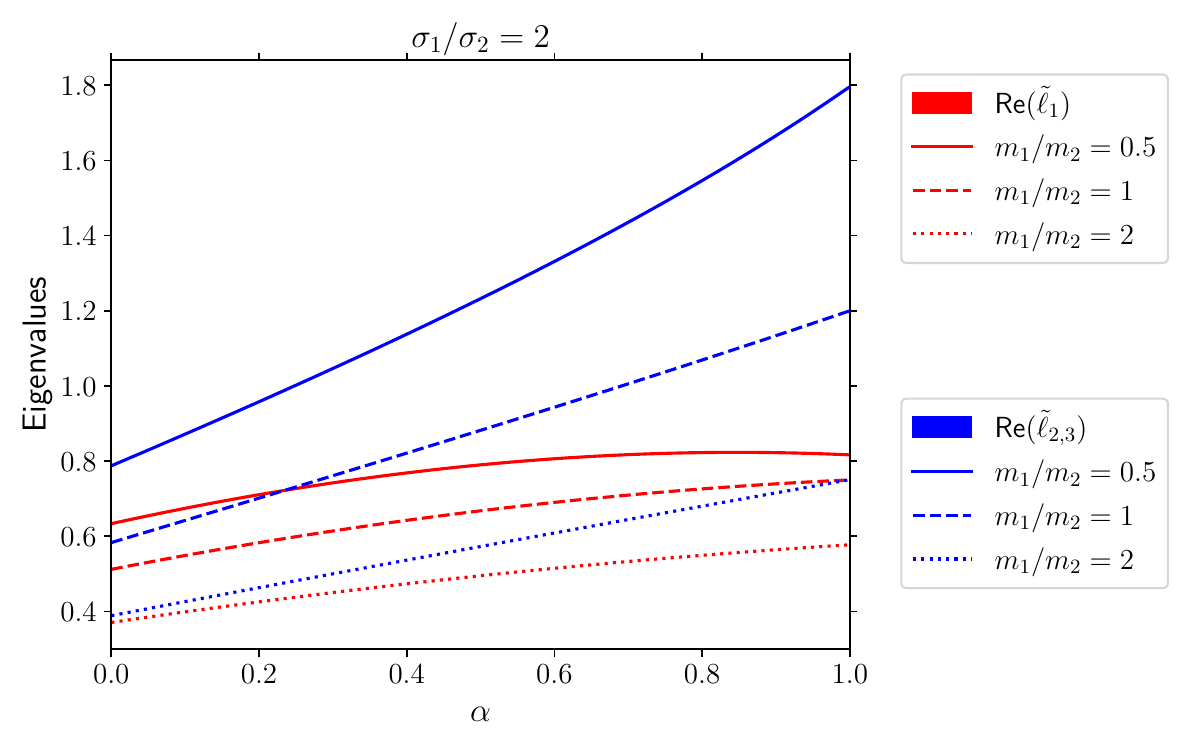}}
\end{tabular}
\end{center}
\caption{Plot of the real parts of the eigenvalues $\widetilde{\ell}_i$ ($i=1,2,3$) as functions of the (common) coefficient of restitution $\al_{22}=\al_{12}\equiv \al$ for the diameter ratio $\sigma_1/\sigma_2=2$ and different values of the
mass ratio $m_1/m_2$.
\label{appAfig2}}
\end{figure}

\section{Behavior of the zeroth-order pressure tensors near the
steady state}
\label{appB}

In this Appendix we give the expressions of the derivatives of the zeroth-order pressure tensors $P_{k\ell}^*$ and $P_{1,k\ell}^*$ with respect to $a^*$ near the steady state. We consider first IMM where the (dimensionless) elements of the pressure tensor $P_{k\ell}^*$ obey the equation
\begin{equation}
\label{b1}
-\left(\frac{2}{d}a^* P_{xy}^{*}+\zeta^{*}\right)\left(1-\frac{1}{2}a^*\frac{\partial}{\partial
a^*}\right) P_{k\ell}^{*}+ a_{k\mu}^*P_{\ell\mu}^{*}+a_{\ell\mu}^*P_{k\mu}^{*}= -\left[\nu_\eta^* P_{k\ell}^{*}+\left(\zeta^*-\nu_\eta^*\right)\delta_{k\ell}\right],
\end{equation}
where in the tracer limit $P_{2,k\ell}^{*}\simeq P_{k\ell}^{*}$ and $\zeta_2^*\simeq \zeta^*$. From Eq.\ \eqref{b1}, one gets the set of equations
\beq
\label{b2}
\frac{\partial P_{yy}^*}{\partial a^*}=\frac{2(\nu_\eta^*-\zeta^*)-2P_{yy}^*\left(\nu_\eta^*-\zeta^*-\frac{2}{d}a^* P_{xy}^*\right)}{a^*\left(\frac{2}{d}a^* P_{xy}^{*}+\zeta^{*}\right)},
\eeq
\beq
\label{b3}
\frac{\partial P_{xy}^*}{\partial a^*}=\frac{-2P_{yy}^*a^*-2P_{xy}^*\left(\nu_\eta^*-\zeta^*-\frac{2}{d}a^* P_{xy}^*\right)}{a^*\left(\frac{2}{d}a^* P_{xy}^{*}+\zeta^{*}\right)}.
\eeq
As expected, the numerators and denominators of equations\ \eqref{b2} and \eqref{b3} vanish in the steady state $[(2/d)a^*P_{xy}^*+\zeta^*=0]$. As in the case of IHS \cite{G07}, the steady-state limit of equations\ \eqref{b2} and \eqref{b3} can be evaluated by means of l'Hopital's rule. In this case, one achieves the results
\beq
\label{b4}
\Upsilon_{yy}=4 P_{yy}^*\frac{a^*\Upsilon_{xy}+P_{xy}^*}{2a^{*2}\Upsilon_{xy}+d\left(2\nu_\eta^*-\zeta^*\right)},
\eeq
where $\Upsilon_{yy}\equiv \left(\partial P_{yy}^*/\partial a^*\right)_s$ and $\Upsilon_{xy}\equiv \left(\partial P_{xy}^*/\partial a^*\right)_s$ is the real root of the cubic equation
\begin{align}
\label{b5}
2 a^{*4} \Upsilon_{xy}^3+4da^{*2}\nu_\eta^*\Upsilon_{xy}^2+\frac{d^2}{2}\left(4\nu_\eta^{*2}+6
\zeta^*\nu_\eta^*-3\zeta^{*2}\right)\Upsilon_{xy}\nonumber\\+d^2\left(\nu_\eta^*-\zeta^*\right)
\nu_\eta^{*-2}\left(\zeta^{*2}-5\zeta^*\nu_\eta^*+
2\nu_\eta^{*2}\right)=0.
\end{align}
In the above equations, it is understood that all the quantities are computed in the steady state.

We consider now the derivatives of the elements $P_{1,k\ell}^*$ with respect to $a^*$. They verify the time dependent equation
\begin{equation}
\label{b6}
-\left(\frac{2}{d}a^* P_{xy}^{*}+\zeta^{*}\right)\left(1-\frac{1}{2}a^*\frac{\partial}{\partial
a^*}\right) P_{1,k\ell}^{*}+ a_{k\mu}^*P_{1,\ell\mu}^{*}+a_{\ell\mu}^*P_{1,k\mu}^{*}= Y \delta_{k\ell}+X_0 P_{1,k\ell}^*+X P_{k\ell}^*,
\end{equation}
where we recall that the quantities $Y$, $X_0$ and $X$ for IMM are defined by equations\ \eqref{3.8} and \eqref{3.9}, respectively. The derivatives of $\partial_{a^*}P_{1,yy}^{*}$, $\partial_{a^*}P_{1,xy}^{*}$, and $\partial_{a^*}\gamma$ can be easily obtained from the results derived for IHS in \cite{G07} by replacing the expressions of the quantities $Y$, $X_0$, and $X$ of IHS by their corresponding counterparts for IMM given by equations\ \eqref{3.8} and \eqref{3.9}, respectively. Thus, the expressions of the derivatives $\partial_{a^*}P_{1,xy}^{*}$ and $\partial_{a^*}P_{1,xy}^{*}$ are given by equations\ (B.16) and (B.17), respectively, of \cite{G07} while the derivative $\partial_{a^*}\gamma$
is\footnote{Some typos were found in Eq.\ (B.22) of \cite{G07} while  the present paper was written. The expressions displayed here are the corrected results}
\beq
\label{b7}
\left(\frac{\partial \gamma}{\partial a^*}\right)_s=\frac{\Lambda_1}{\Lambda_2},
\eeq
where
\begin{align}
\label{b8}
\Lambda_1=d\left(\frac{1}{2}a^*\chi-X_0\right)\Bigg[\left(\frac{1}{2}a^*\chi-X_0\right)
\left(\gamma\chi-\frac{2}{d}P_{1,xy}^*\right)-\frac{2}{d}a^*\Big(\chi P_{1,xy}^*-P_{1,yy}^*+X \Upsilon_{xy}
\Big)\nonumber\\ +2a^{*2}\Big(\chi P_{1,yy}^*+X \Upsilon_{yy}\Big)\Bigg],
\end{align}
\begin{align}
\label{b9}
\Lambda_2=d\left(\frac{1}{2}a^*\chi-X_0\right)\Bigg[\left(\frac{1}{2}a^*\chi-X_0\right)\left(\zeta_1^*+\frac{1}
{2}a^*\chi+\gamma \zeta_1^*\right)+\frac{2}{d}a^*\left(X_0'P_{1,xy}^*+X'P_{xy}^*\right)\Bigg]\nonumber\\ -2a^{*2}\left(Y'+X_0'P_{1,yy}^*+X'P_{yy}^*\right).
\end{align}
In equations\ \eqref{b8} and \eqref{b9}, $\chi=(2/d)(P_{xy}^*+a^* \Upsilon_{yy})$, $Y'=\partial_\gamma Y$, $X'=\partial_\gamma X$, and $X_0'=\partial_\gamma X_0$. As in the case of the excess granular gas, all the quantities appearing in equations\ \eqref{b7}--\eqref{b9} are evaluated in the steady state.

In the case of the VGS kinetic model, the expressions of the derivatives $\Upsilon_{yy}$ and $\Upsilon_{xy}$ are given by equations \ \eqref{b4} and \eqref{b5}, respectively, except that $\nu_\eta^*=(1+\al_{22})/2+\epsilon_{22}^*$ and $\zeta^*$ must be replaced by $\epsilon_{22}^*$. With respect to the derivatives associated with the tracer particles, their forms are identical to those obtained for IMM except that $X=0$, and the quantities $Y$ and $X_0$ are given by
\beq
\label{b10}
Y=\frac{1+\al_{12}}{2}\nu_{12}^* \Big[\gamma+2\mu_{12}\mu_{21}(1-\gamma)
\Big],
\eeq
\beq
\label{b11}
X_0=-\frac{1+\al_{12}}{2}\nu_{12}^*-\epsilon_{12}^*.
\eeq

%

\end{document}